\documentclass[twocolumn,pra,aps,showpacs,superscriptaddress,floatfix,showkeys,nofootinbib,10pt]{revtex4-1}

\usepackage{amsmath,amsfonts,amssymb,amsthm}
\usepackage{color,graphicx}
\usepackage{bbm,bm}
\usepackage{booktabs}
\usepackage{multirow}
\usepackage{bbold}
\usepackage{hyperref}
\usepackage{comment}
\usepackage{xr}

\newtheorem{lemma}{Lemma}
\newtheorem{proposition}{Proposition}

\def\beq{\begin{eqnarray}}
\def\eeq{\end{eqnarray}}
\def\blem{\begin{lemma}}
\def\elem{\end{lemma}}
\def\bprop{\begin{proposition}}
\def\eprop{\end{proposition}}
\def\bprop{\begin{proposition}}
\def\eprop{\end{proposition}}

\newcommand{\tr}{\text{tr}}

\newcommand{\Ps}{\bar{P}_\text{succ}}
\newcommand{\vx}{{\vec{x}}}
\newcommand{\Bl}{\mathcal{B}}
\newcommand{\Ll}{\mathcal{L}}
\newcommand{\Il}{\mathcal{I}}
\newcommand{\Hl}{\mathcal{H}}
\renewcommand{\rho}{\varrho}

\newcommand{\Id}{\mathbb{1}}

\newcommand{\ket}[1]{ | #1 \rangle}
\newcommand{\bra}[1]{ \langle #1 |}
\newcommand{\proj}[1]{\ket{#1}\hspace{-2.5pt}\bra{#1}}
\newcommand{\kb}[2]{\ket{#1}\hspace{-2.5pt}\bra{#2}}
\newcommand{\bk}[2]{\langle #1  |  #2 \rangle}

\newcommand{\norm}[1]{\left|\hspace{-1pt}\left|\hspace{1pt} #1 \hspace{1pt}\right|\hspace{-1pt}\right|}

\begin{document}
	
	\title{Semi-device-independent self-testing of unsharp measurements}
	
	\author{Nikolai Miklin} \email{nikolai.miklin@ug.edu.pl}\affiliation{Institute of Theoretical Physics and Astrophysics, National Quantum Information Center, Faculty of Mathematics, Physics and Informatics, University of Gdansk, 80-952 Gda\'nsk, Poland}
	\author{Jakub J. Borka{\l}a} \affiliation{Institute of Theoretical Physics and Astrophysics, National Quantum Information Center, Faculty of Mathematics, Physics and Informatics, University of Gdansk, 80-952 Gda\'nsk, Poland}
	\author{Marcin Paw{\l}owski} \affiliation{Institute of Theoretical Physics and Astrophysics, National Quantum Information Center, Faculty of Mathematics, Physics and Informatics, University of Gdansk, 80-952 Gda\'nsk, Poland}

	\date{\today}
	
	\begin{abstract}
	Unsharp quantum measurements provide a resource in scenarios where one faces the trade-off between information gain and disturbance. In this work we introduce a prepare-transform-measure scenario in which  two-outcome unsharp measurements outperform their sharp counterparts, as well as any stochastic strategy involving dichotomic projective measurements. Based on that, we propose a scheme for semi-device-independent self-testing of unsharp measurements and show that all two-outcome qubit measurements can be characterized in a robust way. Along with the main result, in this work we introduce a method, based on semidefinite programming, for bounding quantum correlations in scenarios with sequential measurements of length two.~This method can also be applied to refine security analysis of the semi-device-independent one-way quantum key distribution. We also present new information gain--disturbance relation for pairs of dichotomic measurements.     
	    
	\end{abstract}

	\maketitle

\section{Introduction} 
Nonprojective, or generalized, measurements are known to provide an advantage in various quantum information processing tasks. Examples include quantum tomography~\cite{renes2004symmetric,bisio2009optimal}, state discrimination~\cite[]{chefles2000quantum,barnett2009quantum}, and randomness certification~\cite{acin2016optimal}, to name a few. However, there is one important property of generalized measurements that is discussed less often than the others. It is the trade-off between information gain and disturbance that this type of measurement may allow for. 

In quantum measurement, if some information is obtained about the state of the measured system, that state will necessarily be perturbed. This principle, which was formalized in Ref.~\cite[]{busch2009no}, is central to quantum mechanics and it is also one of the pillars of quantum cryptography~\cite[]{bennett2014quantum}. Quantitatively the interplay between information gain and disturbance has been formulated in numerous trade-off inequalities~\cite[]{fuchs1996quantum, fuchs1998information, kretschmann2008information, buscemi2008global}. 

Sharp or ``strong" measurement does not provide such a trade-off. The state of the system after sharp quantum measurement corresponds to one of the eigenstates of the measured observable. It means that the disturbance caused by sharp measurements is maximal and any subsequent measurement on the same system will not provide any additional information about its original state. On the other hand, maximally unsharp measurement, corresponding to an identity channel applied on the state, does not cause any disturbance, but at the same time it does not give any information about the state of the system. In the case of two-outcome qubit measurements, which we discuss in this paper, the above two types of measurements are precisely the two variants of projective measurements. Hence, the aforementioned trade-off is possible only if one has access to a device performing nonprojective, or generalized, measurements. 

At this point we should clarify some terminology that we use in this paper. Throughout the text we use the terms ``two-outcome", ``binary", and ``dichotomic" interchangeably.  For two-outcome qubit measurements the terms ``unsharp" and ``nonprojective" mean the same thing except for the case of maximally unsharp measurements, which are trivial measurements and are not interesting to study. Keeping this in mind, we note that the term ``nonprojective" is very often associated exclusively with three- or more-outcome measurements in the case of qubit systems. Hence, we chose to use the term ``unsharp" to denote two-outcome nonprojective measurement throughout this work. Unsharp measurements that aim at bringing minimal disturbance to the system are often referred to as ``weak measurements"~\cite[]{aharonov1988result}. 

Unsharp and, in particular, weak measurements have been the subject of numerous studies~\cite[]{busch1986unsharp,busch1988surprising,ritchie1991realizaiton, fuchs1996quantum, fuchs1998information, kretschmann2008information, buscemi2008global, pryde2005measurement,aharonov2014foundations}. We would like to mention some of the recent results of Refs.~\cite[]{silva2015multiple,curchod2017unbounded,curchod2018entangled} showing that nondestructive sequential unsharp measurements can be utilized to establish nonlocal correlations among multiple parties from one entangled pair and generate an unbounded amount of certified randomness. Naturally, given their importance, unsharp quantum measurements were reportedly implemented by various experimental groups~\cite[]{ritchie1991realizaiton,pryde2005measurement}. However, all of these results require one's trust in the measuring or preparation devices. 

In this paper we ask the following questions: Is it possible to certify implementation of an unsharp measurement without making any assumptions, or with minimal assumptions made about the inner workings of the devices? If so, can one characterize the performed measurement based solely on the observed statistics? The answer comes from the rapidly growing field of \emph{self-testing}~\cite[]{mayers2003self,vsupic2019self}. It turns out that in quantum experiments it is possible to certify implementation of the target measurement or preparation of the desired state from the statistics alone and with minimal assumptions made about the inner workings of the devices. The most prominent example is self-testing of a two-qubit Bell state in the case of maximal violation of Clauser-Horne-Shimony-Holt (CHSH) inequality~\cite[]{clauser1969proposed,braunstein1992maximal,popescu1992states}.       
  
Self-testing was originally proposed in the framework of the Bell test~\cite[]{mayers2003self}, also known as device-independent (DI) characterization. However, it was soon generalized to a more experimentally appealing prepare-and-measure scenario~\cite[]{tavakoli2018self}, which followed the ideas of semi-device-independent (SDI) characterization of quantum systems~\cite[]{pawlowski2011semi}. In the SDI framework one assumes that the dimension of the degree of freedom of a physical system used for information processing is bounded from above. We would like to point out that the assumption on the dimension is not only natural for studying generalized measurements (due to the possibility of a dilation), but it is also necessary for obtaining nontrivial results in scenarios with sequential measurements, as widely discussed in the literature~\cite[]{fritz2010quantum,hoffmann2018structure,budroni2019memory}. Additionally, more and more works on self-testing are now considering the SDI framework~\cite[]{mironowicz2018experimentally,tavakoli2018self,tavakoli2018self2,farkas2019self}.

The majority of self-testing results, in both DI and SDI settings, consider target objects, e.g.,~an entangled state, that satisfy the property of extremality. In this case, one can hope to find a witness, e.g.,~a Bell inequality, that singles it out. In our case, the situation is somewhat more complicated. The reason is that statistics of an unsharp binary qubit measurement can be reproduced exactly by a statistical mixture of projective measurements on the same state space without any postprocessing~\cite[]{oszmaniec2017simulating}. As a consequence, if one considers a simple prepare-and-measure scenario, it will not be possible to distinguish whether the measuring device performs an actual unsharp measurement or its simulation.

In this work we overcome the above limitations by considering a tripartite prepare-transform-measure scenario. We introduce a game in which unsharp measurements outperform their sharp counterparts as well as any statistical mixture of projective measurements. This claim is formalized by two propositions in Section~\ref{sec:resource}. In Section~\ref{sec:selftest} we show that in the proposed scenario one can device-independently characterize all two-outcome unsharp qubit measurements, subject to an assumption that the physical system that parties use for information processing is a qubit. In the same section we show that this characterization is robust to noise. In our proofs regarding self-testing we use semidefinite-programming techniques that we describe in Section~\ref{sec:method}. These techniques can be used to tackle many other problems involving sequential measurements outside of the current work. We demonstrate it by refining the security analysis of the SDI one-way quantum key distribution~\cite[]{pawlowski2011semi} in Section~\ref{sec:appl}. Some proofs as well as minor technical details are left for the Appendix. 

	\section{Prepare-transform-measure scenario} 
\begin{figure}[t!] \centering
		\includegraphics[width=.3\textwidth]{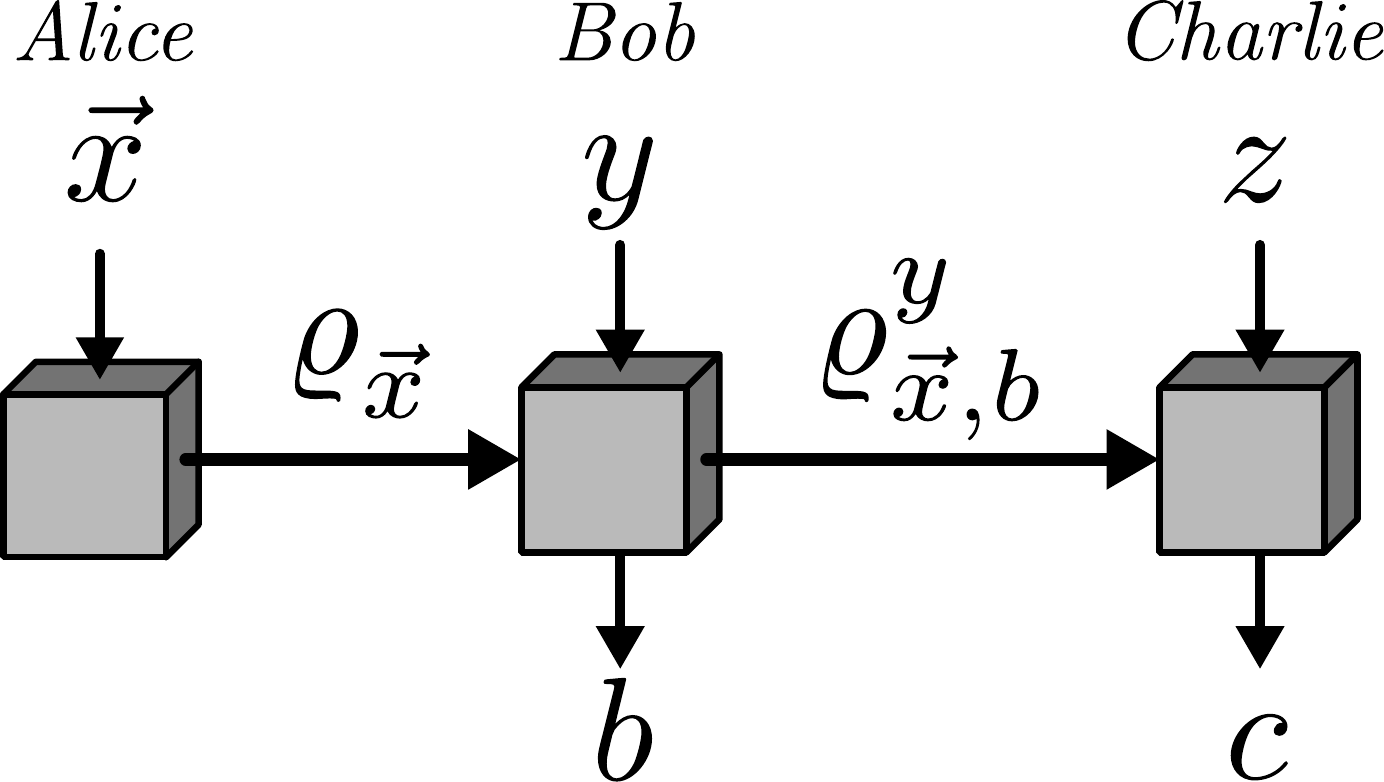}
		\caption{Scenario: $\vec{x}$, input data of Alice; $y$ and $b$, input and output of Bob; $z$ and $c$, input and output of Charlie; $\rho_\vx$, states that Alice sends to Bob; and $\rho^{y}_{\vx,b}$, states that Bob sends to Charlie.}
		\label{fig:scenario}
	\end{figure}

	Let us now introduce our scenario which is a generalization of $2\to{1}$ quantum random access code (QRAC)~\cite[]{ambainis2008quantum}. Consider three parties, Alice, Bob, and Charlie, communicating in a sequential way as shown in Fig.~\ref{fig:scenario}. Alice receives two random bits $\vec{x} = (x_0,x_1),\; x_0,x_1 \in \{0,1\}$. Bob and Charlie each receive a random bit, $y$ and $z$ respectively, which indicates the bit of Alice this party needs to guess. All the random bits $x_0,x_1,y$, and $z$ are independent and uniformly distributed. We assume that the parties do not share any entangled states, however, they may have access to any amount of shared classical randomness. Depending on $\vx$, Alice prepares a qubit state $\rho_\vx$ which she sends to Bob. Bob performs some quantum operation on $\rho_\vx$, depending on his input $y$, and obtains a classical outcome $b$. Afterwards, Bob sends the postoperation state $\rho_{\vx,b}^{y}$ to Charlie, who performs a measurement depending on $z$. 
	
A figure of merit that we consider is the following average success probability: 
\begin{equation}
\label{eq:f_merit}
\Ps = \frac{\alpha}{8}\sum\limits_{\vx,y} \text{Pr}(b=x_y|\vx,y) + \frac{1-\alpha}{8}\sum\limits_{\vx,z} \text{Pr}(c=x_z|\vx,z),
\end{equation}
with $\alpha\in [0,1]$ and where summation is taken over all values of $\vx,y$ and $\vx,z$, respectively. The parameter $\alpha$ is announced prior to the game and remains unchanged throughout all the rounds. It dictates the parties whose guess contributes more to the overall success probability. As we show later in the text, the optimal operations that Bob performs in the case of $\alpha \in (0,1)$ correspond to unsharp measurements.	
	
\subsection{Preliminary notions}
Measurements in quantum mechanics are formalized by the notion of positive-operator valued measure (POVM). For two-outcome measurements, which we consider in this paper, a POVM is an ordered pair of positive semidefinite linear operators $(B_0,B_1)$, $B_i\in \Ll(\Hl), B_i\geq 0, i=0,1$, satisfying $B_0+B_1 = \Id$, where $\Id$ is the identity operator. Projection-valued measure (PVM) is a special case that corresponds to $B_i^2 = B_i, i=0,1$. From here it is clear that the only two possibilities of two-outcome qubit PVMs are sharp measurements, with both $B_i$ being normalized rank-$1$ operators, and the maximally unsharp, or trivial, measurement with one $B_i$ equal to $\Id$ and the other being zero. We call a POVM $(B_0,B_1)$ \emph{unbiased} if $\tr(B_0) = \tr(B_1) = 1$ and biased otherwise. In this work we concentrate on the former case as it encapsulates the resources that are interesting to us for the trade-off between information gain and disturbance. The biased case allows for an additional flexibility with respect to probability distribution of the measurement outcomes.
	
In order to describe the operations of Bob we need to introduce the notion of a quantum instrument, which captures both measurement statistics and the state evolution (see, e.g.,~Ref.~\cite{heinosaari2011mathematical}). By \emph{quantum instrument} for the case of two measurement outcomes $\{0,1\}$ we mean an ordered pair of completely positive trace nonincreasing maps that sum to a channel, completely positive trace-preserving map. Let us denote the instruments of Bob by $(\Bl^{y}_0,\Bl^{y}_1)$ for each $y$, where each $\Bl^{y}_b: \Ll(\Hl) \mapsto \Ll(\Hl)$ and $\dim(\Hl) = 2$, in agreement with our SDI assumption. If we apply these maps to the state $\rho_\vx$ (for a given $\vx$), which Alice sends to Bob, we obtain a pair of positive operators $\sigma^{y}_{\vx,b} = \Bl^{y}_b(\rho_\vx)$, $b=0,1$. Their traces form the probability distribution of Bob's measurement outcomes $p(b|\vx,y) = \tr(\sigma^{y}_{\vx,b})$, and, if normalized, these operators correspond to the postmeasurement states that Bob sends to Charlie, $\rho^{y}_{\vx,b} = \frac{\sigma^{y}_{\vx,b}}{p(b|\vx,y)}$. We say that instrument $(\Bl^{y}_0,\Bl^{y}_1)$ is compatible with a POVM $(B^{y}_0,B^{y}_1)$ if for all states $\rho$, $\tr(\rho B^y_b) = \tr(\Bl^{y}_b(\rho)), b=0,1$. 

Using the formalism of quantum instruments we can write the joint probability distribution of both parties' outcomes as  follows: 
\beq
\label{eq:p_distr_pre}
p(b,c|\vx,y,z) = \tr(\Bl^{y}_b(\rho_\vx)C^{z}_c),
\eeq
where we denoted Charlie's POVMs by $(C^{z}_0,C^{z}_1)$, $z=0,1$. Naturally, using this formula we can calculate the probabilities in our figure of merit in Eq.~(\ref{eq:f_merit}).

The action of an instrument on a state can be specified by its Kraus decomposition (see, e.g.,~Ref.~\cite{heinosaari2011mathematical}). However, it will be more convenient for us to work with instruments in their Choi-Jamio\l{}kowski (CJ) representation (see, e.g.,~Ref.~\cite{heinosaari2011mathematical}). We obtain the CJ operators of Bob's instruments by the following map: 
\beq
\mathfrak{B}^y_b = \left(\Il\otimes\Bl_b^y(\proj{\Omega})\right),
\eeq
where $\Il$ is an identity map on elements of $\Ll(\Hl)$ and $\ket{\Omega}$ is a non-normalized Bell state $\ket{\Omega}=\sum_{i=0,1}\ket{ii}$. The inverse map is given by the relation $\Bl^y_b(\rho) = \tr_1(\rho^T\otimes\Id \mathfrak{B}^y_b)$, where the partial trace is taken with respect to the first subsystem, and $(\cdot)^T$ stands for matrix transpose with respect to the basis $\{\ket{i}\}_{i=0,1}$ in which the CJ isomorphism is defined. From this relation it is clear that POVM $(B^y_0,B^y_1)$, associated with the CJ operator $(\mathfrak{B}^y_0,\mathfrak{B}^y_1)$, is simply a pair of reduced operators with the trace taken over the second (output) subsystem, i.e.,~$B^y_b = \tr_2(\mathfrak{B}^y_b)^T, b=0,1$, for each $y=0,1$.

Using CJ representation of the instruments we can rewrite the conditional probabilities observed in our scenario as follows:
\beq
\label{eq:p_distr}
p(b,c|\vx,y,z) = \tr(\rho^T_\vx\otimes C^z_c\; \mathfrak{B}^y_b).
\eeq
One can immediately notice that the above formula is very similar to the one of tensor product measurements on an entangled state. This motivated us to try to apply the semidefinite programming (SDP) techniques of Ref.~\cite{navascues2015bounding} to bound correlations in Eq.~\eqref{eq:p_distr}. As discussed in Section~\ref{sec:method}, these SDP techniques had to be modified for this purpose.

\section{Unsharpness as a resource}\label{sec:resource}
We have already mentioned in the Introduction that unsharp measurements are needed to access the trade-off between information gain and disturbance. In this section we show that the figure of merit in Eq.~(\ref{eq:f_merit}) allows for witnessing unsharpness of the performed measurement in an SDI way, i.e.,~with no assumptions about the measuring or preparation devices apart from the upper bound on the dimension of the underlined Hilbert space. We are ready to present our first part of the results, which is given by the following two propositions.
\bprop
\label{prop:pvm_bound}
The average success probability $\Ps$ for strategies involving only projective measurements and their probabilistic mixtures is bounded by the following expression:
\beq
\Ps^{\text{PVM}}(\alpha) = \left\{ 
\begin{array}{ll}
      \frac{1}{2}+\frac{1-\alpha}{2\sqrt{2}}, & 0\leq\alpha\leq 1-\frac{2}{\sqrt{7}}  \\
      \frac{1}{2}+\frac{1}{8}\sqrt{4+(1-\alpha)^2}, & 1-\frac{2}{\sqrt{7}}< \alpha\leq \frac{1}{3} \\
      \frac{1}{2}+\frac{1}{4}\sqrt{1+\alpha^2}, & \frac{1}{3}< \alpha\leq 1. 
\end{array} \right. \nonumber
\eeq
\eprop

\begin{figure}[t!]\centering
		\includegraphics[width=.45\textwidth]{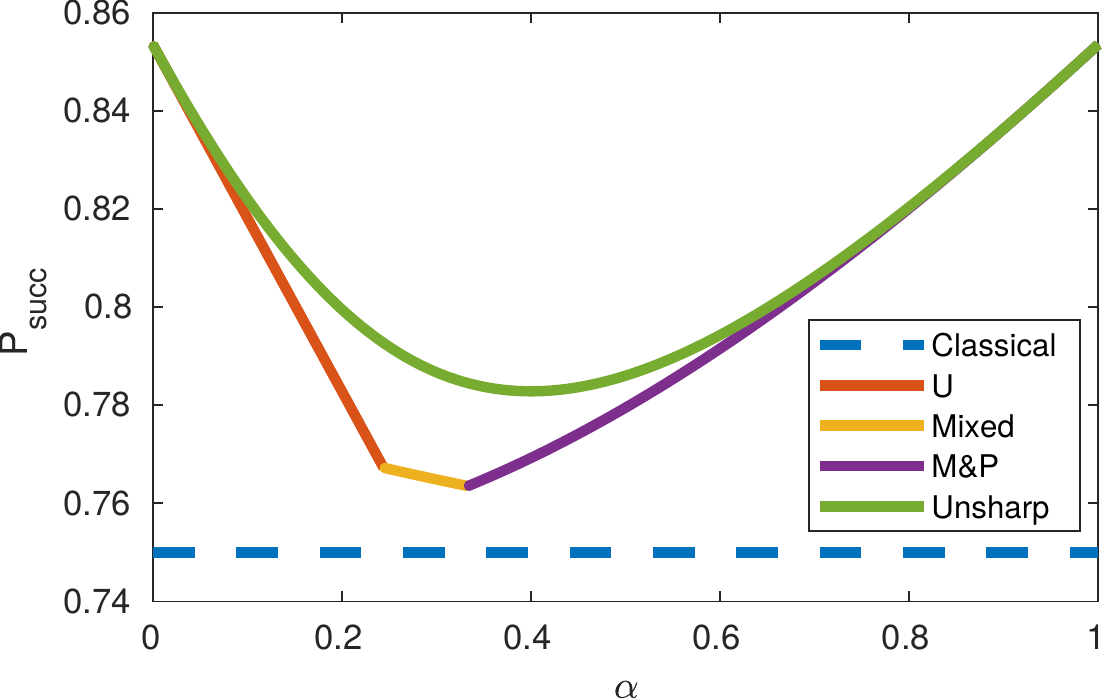}
		\caption{Bounds on the average success probability. Blue (dashed) line corresponds to the classical strategy. The three-segment line corresponds (from left to right) to the unitary (orange), mixed (yellow) and measure and prepare (purple) projective strategies, respectively. The green (uppermost) line corresponds to the general strategy with unsharp measurements. All the bounds are tight.}
		\label{fig:bounds}
\end{figure}

The reason that the function $\Ps^{\text{PVM}}(\alpha)$ has three distinct segments is that Bob can choose to apply one of the three different optimal strategies depending on $\alpha$. The first one, which we call ``unitary", corresponds to the maximally unsharp measurement $(\Id,0)$ for both $y=0,1$, which is compatible with Bob outputting $b=0$ and applying a unitary channel on Alice's states. The third one is the ``measure-and-prepare" strategy, in which Bob performs a sharp measurement for both $y=0,1$ and prepares some state that he sends to Charlie. The middle segment corresponds to Bob applying the unitary strategy for $y=0$, and the prepare and measure strategy for $y=1$ (or the other way around), which we refer to as a ``mixed" strategy. The proof of Proposition~\ref{prop:pvm_bound} is too lengthy to be included in the main text and is given in Appendix~\ref{app:pvm_bound}.

\bprop
\label{prop:povm_bound}
The bound on the average success probability $\Ps$ for the general strategy involving unsharp measurements is the following
\beq
\Ps^{\text{POVM}}(\alpha) = \frac{1}{2}+\frac{1-\alpha}{4\sqrt{2}}+\frac{1}{4\sqrt{2}}\sqrt{(1-\alpha)^2+4\alpha^2}.\nonumber
\eeq

\eprop
The results of both Propositions~\ref{prop:pvm_bound}, and \ref{prop:povm_bound} are shown in Figure~\ref{fig:bounds}. In this figure we also show the classical bound for this game, which evidently equals $\frac{3}{4}$ irrespective of $\alpha$. We should point out that all the presented bounds are tight in the sense that there exist states and measurements reaching these average success probabilities. 

The optimal quantum strategy that attains the bound in Proposition~\ref{prop:povm_bound} is the following. The states of Alice have the same form as in the $2\to 1$ QRAC, i.e.,~independent of $\alpha$. Their explicit form is given below, where we denote $\rho^T_\vx = \proj{\psi_\vx}$:
\beq
\label{eq:expl_states}
&&\ket{\psi_{00}} = \frac{\ket{0}+\ket{+}}{\sqrt{2+\sqrt{2}}}, \quad \ket{\psi_{01}} = \frac{\ket{0}+\ket{-}}{\sqrt{2+\sqrt{2}}},\nonumber \\
&&\ket{\psi_{10}} = \frac{\ket{1}+\ket{+}}{\sqrt{2+\sqrt{2}}}, \quad \ket{\psi_{11}} = \frac{\ket{1}-\ket{-}}{\sqrt{2+\sqrt{2}}}.
\eeq 
The optimal POVMs of Charlie can be set to be the following:
\beq
&&(C^0_0,C^0_1) = (\proj{0},\proj{1})\\ 
&&(C^1_0,C^1_1) = (\proj{+},\proj{-})\nonumber,
\eeq
and the optimal instruments of Bob (their CJ operators) can be taken to be of the form $\mathfrak{B}^y_b = \proj{\beta^y_b}$,
\beq
\label{eq:intr_cj}
&&\ket{\beta^0_0} = \sqrt{\lambda}\ket{00}+\sqrt{1-\lambda}\ket{11},\nonumber\\
&&\ket{\beta^0_1} = \sqrt{1-\lambda}\ket{00}+\sqrt{\lambda}\ket{11},\\
&&\ket{\beta^1_0} = \sqrt{\lambda}\ket{++}+\sqrt{1-\lambda}\ket{--},\nonumber\\
&&\ket{\beta^1_1} = \sqrt{1-\lambda}\ket{++}+\sqrt{\lambda}\ket{--},\nonumber
\eeq
where we left the real parameter $\lambda$ unspecified. It is clear that $\lambda$ determines the sharpness of the corresponding measurement of Bob. 
We find that the optimal $\lambda$ is the following function of $\alpha$:
\beq
\label{eq:opt_norm_app}
\lambda = \frac{1}{2}+\frac{\alpha}{\sqrt{(1-\alpha)^2+4\alpha^2}},
\eeq
which leads to the expression of the optimal average success probability from Proposition~\ref{prop:povm_bound}. 

The above analysis gives only a lower bound on the average success probability. In order to prove that this bound is also an upper bound we used the aforementioned SDP techniques, described in Section~\ref{sec:method}. We compared the numerical results from the SDP for each value of $\alpha$ with a step of $0.001$ with the lower bound values from Proposition~\ref{prop:povm_bound} and the discrepancy was always less than or of the order of $10^{-9}$, which leads us to conclude that the obtained bound is in fact the correct one. 

Having a gap between success probabilities achieved by projective measurements and unsharp measurements is necessary for the self-testing claims. However, already from the results of Propositions~\ref{prop:pvm_bound} and~\ref{prop:povm_bound} it follows that if one experimentally obtains a value of average success probability that is greater than $\Ps^{\text{PVM}}(\alpha)$ for some $\alpha$, one can conclude that Bob's measuring device performed an unsharp measurement. In order to answer the question of exactly which measurements was performed by Bob we will need a stronger result, given in the next section.

\section{SDI characterization of unsharp measurements}\label{sec:selftest} 
The core idea that we use in our paper is to certify implementation of a quantum instrument that realizes the target unsharp measurement. The set of quantum instruments realizing a given unsharp POVM is strictly larger than the set of probabilistic mixtures of instruments corresponding to PVMs that reproduce the same measurement statistics for Bob. It is surprising that a simple and intuitive figure of merit in Eq.~(\ref{eq:f_merit}) is sufficient to witness this fact. Although the target measurements that we would like to self-test are not extremal in the space of all POVMs, some instruments that realize these measurements are extremal and hence can potentially be witnessed by saturating the bound in Proposition~\ref{prop:povm_bound}.

In this paper we focus on certification of the unsharpness parameter of the performed measurement from the observed data. In this section we only consider two-outcome unbiased qubit POVMs. Hence, the unsharpness parameter, which can be taken to be the operator norm of one of the effects completely characterizes these measurements up to an isometry. Hence, this SDI characterization is equivalent to the SDI self-testing of individual measurements. 

Since the figure of merit in Eq.~(\ref{eq:f_merit}), when written in terms of probabilities in Eq.~(\ref{eq:p_distr}), is linear with respect to CJ operators of Bob's instruments, it is sufficient to consider only extremal instruments. In Appendix~\ref{app:instr_rank_1} we show that in our case the extremal instruments have the form as in Eq.~(\ref{eq:intr_cj}) (with $\lambda$ being dependent on $y$, in general). Hence, the sharpness of the corresponding POVM is exactly the parameter $\lambda$ in Eq.~(\ref{eq:opt_norm_app}).  

Now, we come to our main question of what can be said about the unsharpness parameter $\lambda$ for each $y$ given the observed value of the average success probability in Eq.~(\ref{eq:f_merit}). We answer this question by setting the opposite problem: What are the maximal values of the average success probability attainable by POVMs $({B}^y_0,{B}^y_1)$ with a given norm? It turns out that the optimal $({B}^y_0,{B}^y_1)$ is always unbiased for the figure of merit in Eq.~(\ref{eq:f_merit}). 

\begin{figure}[t!]\centering
		\includegraphics[width=.45\textwidth]{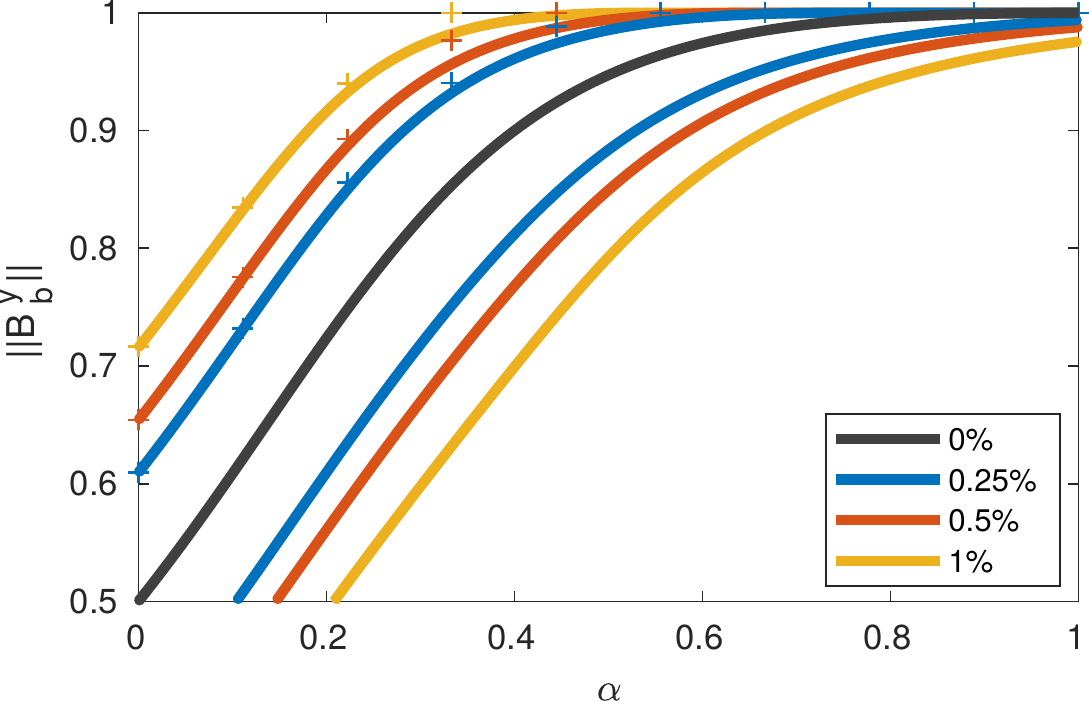}
		\caption{Illustration of the results on robust SDI characterization of unsharp measurements. The optimal operator norm (black line) and bounds on the norm  (color or grey lines) for different values of deviation from the optimal success probability. The crosses represent the discrepancy between the solution which we believe is optimal and the bound that we could prove with our SDP techniques. For more details please see Appendix~\ref{app:robustness}.}
		\label{fig:norm}
\end{figure}

Our results on the robust SDI characterization of unsharp measurements are shown in Figure~\ref{fig:norm}. The black line represents the case of the ideal statistics, in which case the norm of the target measurements $({B}^y_0,{B}^y_1)$ is the same for both $y=0,1$ and is given by the expression in Eq.~(\ref{eq:opt_norm_app}). As we can see, the above norm is a monotonically increasing function of $\alpha$ and its image is the interval $[\frac{1}{2},1]$. This fact suggests that we can self-test any unsharp unbiased two-outcome qubit measurement by picking the corresponding value of $\alpha$. 

In Fig.~\ref{fig:norm} each pair of lines of the same color represents upper and lower bounds on the norm of the POVM for each $y=0,1$, given that the average success probability is above some fixed value. We would like to stress that we do not assume that both norms for $y=0$ and $y=1$ are equal. In Figure~\ref{fig:norm} we show four values $\{0\%, 0.25\%, 0.5\%, 1\%\}$ for $\frac{\varepsilon}{\Ps^{\text{POVM}}(\alpha)}$, where $\epsilon$ is the deviation of the observed value of $\Ps$ from the optimal value $\Ps^{\text{POVM}(\alpha)}$. From the plot one can see that for $\varepsilon$ going to $0$, the region of the compatible norms shrinks monotonically to a point with $\lambda$ given by Eq.~(\ref{eq:opt_norm_app}) for both $y$. This fact suggests that our SDI characterization is indeed robust.    

We would like to give an example of how to read the plots in Fig.~\ref{fig:norm}. Let us say we wish to self-test a measurement with the norm equal to $\frac{1}{2}+\frac{1}{2\sqrt{5}}$. We would pick $\alpha=0.2$, for which POVMs with this norm are the optimal ones. If now our observed average success probability is equal to $0.7975$, which is $0.25\%$ smaller than the optimal value $\Ps^{\text{POVM}}(0.2) \approx 0.7995$, we can conclude that the norm of the implemented POVM for both $y=0,1$ must lay within the interval $[0.6095, 0.8281]$. A more detailed description of the robustness analysis is given in Appendix~\ref{app:robustness}. In Appendix~\ref{app:biased} we also give a sufficient numerical evidence that our scheme can be adapted to SDI characterization of biased unsharp measurements. 

\section{Methods}\label{sec:method}
The proof of Proposition~\ref{prop:pvm_bound} is almost entirely analytical (see Appendix~\ref{app:pvm_bound}). For the proof of Proposition~\ref{prop:povm_bound} as well as for the results of the previous section we used SDP techniques that we describe in this section. These SDP techniques are based on the hierarchy of Navascu{\'e}s and V{\'e}rtesi~\cite[]{navascues2015bounding}. Below, we also provide an argument why the hierarchy of Ref.~\cite[]{navascues2015bounding} cannot be directly applied to our problem. Moreover, our SDP techniques can be applied to a variety of problems involving two sequential measurements on a single system. For example, in Section~\ref{sec:appl} we refine the analysis of the SDI one-way quantum key distribution proposed in Ref.~\cite[]{pawlowski2011semi}. There we show that a secure key can be distilled whenever the success probability of a QRAC exceeds the classical limit of $\frac{3}{4}$, which is a significant improvement compared to the original result~\cite[]{pawlowski2011semi}.   

The SDP hierarchy of Navascu{\'e}s and V{\'e}rtesi~\cite{navascues2015bounding} is very useful for bounding correlations in the Bell scenario with an additional restriction on the local dimension. One can also use this hierarchy to approximate the sets of quantum correlations in the prepare-and-measure scenario, where the dimension constraint is crucial. In particular, using Ref.~\cite{navascues2015bounding} one can directly obtain upper bounds on success probabilities of QRACs. We do not summarize the method of Ref.~\cite{navascues2015bounding} here, so the interested reader should first make themselves familiar with it before reading the following description of the proposed modification.   

In our modification of the hierarchy of Ref.~\cite{navascues2015bounding}, which we formulate here for what is know as the ``1+AB" level, we consider the following set of operators 
\beq
\mathcal{O} = \{\Id\otimes \Id\}\cup\{\rho_\vx\otimes \Id\}_\vx\cup\{\Id\otimes C^z_0\}_{z}, \{\rho_\vx\otimes C^z_0\}_{\vx,z}.\nonumber
\eeq
We then consider four moment matrices $\Gamma^y_b$ for each $y,b$. For each $\Gamma^y_b$ we assume the measure $\tr(\;\cdot\; \mathfrak{B}^y_b)$; i.e., the entries $\Gamma^y_b(O_1,O_2)$ of the matrix correspond to $\tr(O_1^\dagger O_2\mathfrak{B}^y_b)$, $O_1,O_2 \in \mathcal{O}$. It is clear that each $\Gamma^y_b$ must be positive semidefinite, i.e.,~$\Gamma^y_b\geq 0$, $y=0,1,b=0,1$. Also, we can use the main idea of the hierarchy of Ref.~\cite{navascues2015bounding} and construct a basis for $\Gamma^y_b$ from randomly sampled operators of fixed dimension. However, we need to apply more constraints on $\Gamma^y_b$ in order to obtain nontrivial bounds on our figure of merit in Eq.~(\ref{eq:f_merit}). 

In particular, we exploit the fact that $\tr_2(\mathfrak{B}^y_0) + \tr_2(\mathfrak{B}^y_1) = \Id$, for each $y$. From this we can derive the following set of constrains that has to be satisfied by all quantum probabilities $p(b,c|\vx,y,z)$
\beq
\label{eq:sdp}
&&\sum_{b=0,1}\Gamma_b^y(\Id\otimes\Id,\Id\otimes\Id) = 2,\; y=0,1,\\
&& \sum_{b=0,1}\Gamma_b^y(\rho_\vx\otimes\Id,\Id\otimes\Id) = 1,\; \forall \vx, y=0,1,\nonumber\\
&& \sum_{b=0,1}\Gamma_b^0(\rho_\vx\otimes\Id,\rho_{\vec{x'}}\otimes\Id) = \sum_{b=0,1}\Gamma_b^1(\rho_\vx\otimes\Id,\rho_{\vec{x'}}\otimes\Id),\; \forall \vx,\vec{x'}.\nonumber
\eeq
The above constraints should be interpreted as follows. The first type fixes the normalization of the instruments, and the second fixes the normalization of states $\{\rho_\vx\}_\vx$. Finally, the third type of constraint ensures that all the inner products between the states $\{\rho_\vx\}_\vx$ are the same for both choices of Bob's operations. Although the above modification is formulated for our scenario with binary $x_0,x_1,y,z,b$, and $c$, it is straightforward to generalize it to other scenarios with an arbitrary number of settings and outputs. 

Now, we would like to argue why the original hierarchy of Ref.~\cite{navascues2015bounding}, as it stands, could not be applied to our problem. One could, in principle, try to explicitly include the CJ operators $\{\mathfrak{B}^y_b\}_{y,b}$ to the set of operators $\mathcal{O}$ and consider a single moment matrix. However, this would require $\mathfrak{B}^y_b$ to be projectors in some potentially larger Hilbert space, which is not in general the case, since we do not know the normalization of the CJ operator of each individual element of the instrument.

In this paper we have successfully used the proposed SDP method to confirm the bounds obtained with the inner approximation methods. It would be interesting to investigate the convergence of this modified hierarchy~\cite{navascues2015characterizing}.

\section{Applications}\label{sec:appl}
\subsection{Trade-off relation} As an application of the results of Proposition~\ref{prop:povm_bound}, we can derive new trade-off relations in terms of QRAC success probabilities attainable in the described sequential scenario. Indeed, our figure of merit can be rewritten as $\Ps = \alpha \bar{P}_{\text{Bob}} + (1-\alpha) \bar{P}_{\text{Charlie}}$. From the condition $\Ps\leq \Ps^{\text{PVM}}(\alpha)$ we can derive the following trade-off relation 
\beq
\bar{P}_{\text{Charlie}} \leq \frac{1}{2} + \frac{1}{4\sqrt{2}} + \frac{\sqrt{16\bar{P}_{\text{Bob}}(1-\bar{P}_{\text{Bob}})-2}}{8}.\nonumber
\eeq
This relation can be used to calculate the disturbance caused by the measurement of Bob by subtracting $\bar{P}_{\text{Charlie}}$ from the maximally attainable value of $\frac{1}{2}+\frac{1}{2\sqrt{2}}$.

\subsection{Security of the SDI one-way quantum key distribution}
 
Here we demonstrate that the SDP techniques introduced in the previous section have applications outside of the current work. To do this we refine the analysis of the security in semi-device-independent one-way quantum key distribution (QKD), proposed in Ref.~\cite{pawlowski2011semi}. 

\begin{figure}[t!] \centering
		\includegraphics[width=.45\textwidth]{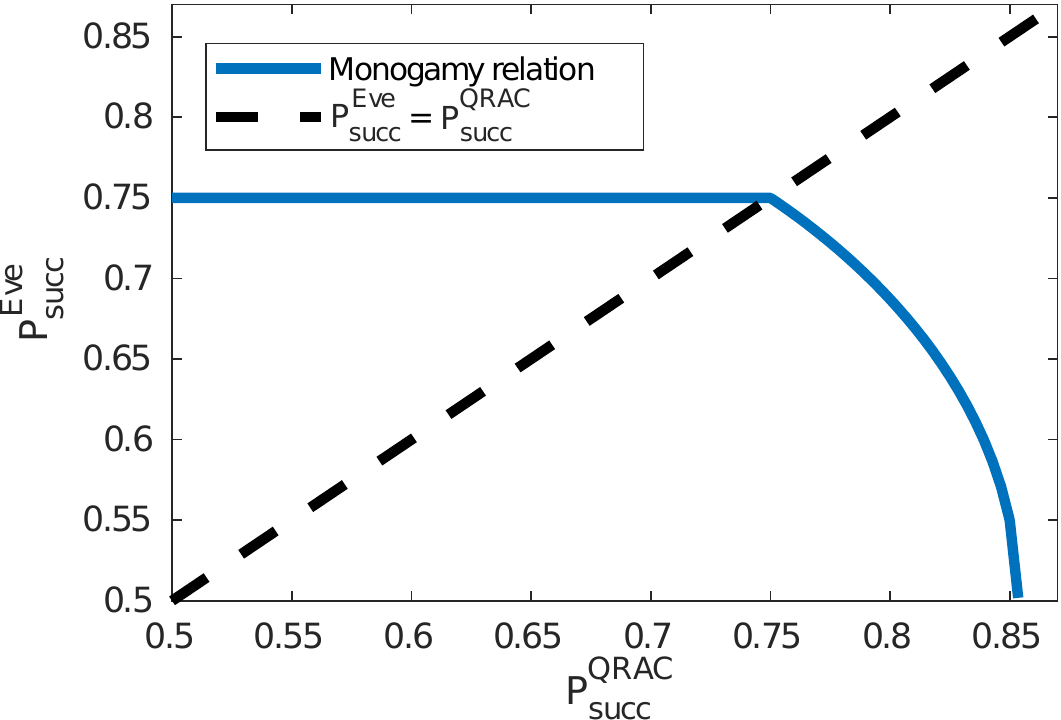}
		\caption{Solid line: monogamy relation for the SDI QKD scheme. Dashed line: the critical regime in which the guessing probabilities of the receiver and the eavesdropper are the same.}
		\label{fig:qkd}
	\end{figure}

Let us briefly introduce this QKD scheme. In this scheme two communicating parties, Alice and Bob are trying to establish a secret key without trusting their devices. As a measure of privacy of the generated key they use the figure of merit of the $2\to 1$ QRAC, which reads
\beq
\Ps^{\text{QRAC}} = \frac{1}{8}\sum_{\vx,y}\text{P}(b=x_y\,|\,\vx,y). 
\eeq
The eavesdropping party, Eve, is trying to guess the bit of Alice corresponding to Bob's input bit $y$, i.e.~to maximize the probability $\Ps^{\text{Eve}} = \frac{1}{8}\sum_{\vx,y}\text{P}(e=x_y\,|\,\vx,y)$, where $e$ is Eve's output. We can then introduce a real parameter $\alpha \in [0,1]$ just like we did in the main text and consider the following objective function
\beq
\Ps = \frac{\alpha}{8}\sum_{\vx,y}\text{P}(b=x_y\,|\,\vx,y) + \frac{1-\alpha}{8}\sum_{\vx,y}\text{P}(e=x_y\,|\,\vx,y),\nonumber \\
\label{eq:f_merit_qkd}
\eeq
which resembles the figure of merit in Eq.~(\ref{eq:f_merit}). We then optimize the expression in Eq.~(\ref{eq:f_merit_qkd}) subject to the constraints of the SDP method described in Section~\ref{sec:method}. Afterwards, by optimizing over $\alpha$ we obtain the monogamy relations between $\Ps^{\text{Eve}}$ and $\Ps^{\text{QRAC}}$ that we plot in Fig.~\ref{fig:qkd}.

As shown in Ref.~\cite{pawlowski2011semi}, Alice and Bob are able to distill a secret key from their QRAC correlations whenever $P^{\text{Charlie}}_{\text{QRAC}}>P^{\text{Bob}}_{\text{Eve}}$. From this condition one can  calculate the critical success probability of QRAC required for the task. In Ref.~\cite{pawlowski2011semi} the authors estimated this critical probability to be $\frac{5+\sqrt{3}}{8}\simeq 0.8415$, which is rather close to the maximal attainable success probability $\frac{1}{2}+\frac{1}{2\sqrt{2}}\simeq 0.8535$ of the $2\to 1$ QRAC. From our plot in Fig.~\ref{fig:qkd} one can see that it is enough for the average success probability $\Ps^{\text{QRAC}}$ to be just above the classical limit of $\frac{3}{4}$.  The latter value was also reported in~Ref.\cite[]{chaturvedi2018security} where the security of the SDI QKD scheme was analyzed from the point of view of the critical detection efficiency of the receiver's detectors. However, to the best of our knowledge, the monogamy relation in Fig.~\ref{fig:qkd} has not appeared in the literature prior to this work.

It is clear that our SDP techniques can be used for the analysis of the security of one-way quantum key distribution in scenarios beyond the one in Ref.~\cite[]{pawlowski2011semi}. Furthermore, there is a way to extend the SDP techniques to analyze correlations in scenarios with sequential measurements of length three. 

\section{Conclusions} In this paper we have discussed the advantages that unsharp measurements can provide if one faces the trade-off between information gain and state disturbance. We have proposed a scheme in which this advantage can be manifested within the semi-device-independent framework. Using this scheme we have shown that all unsharp binary measurements on a qubit can be semi-device-independently characterized in a robust manner, which also implies SDI self-testing of individual measurements for the case of unbiased POVMs. This was possible due to certification of the instruments implementing the respective unsharp measurement. 

In this work we considered only SDI characterization of individual two-outcome qubit measurements since the latter can be achieved by certifying a single real parameter, the unsharpness. However, it is clear that our analysis can be extended to more complicated cases. One generalization, which is definitely worth considering, is certification of trinary qubit measurements. We believe that our scheme can be easily adapted to these purposes.

Semidefinite-programming techniques proposed in this paper can be readily applied to analyze the security of more complex SDI quantum key distribution schemes. These techniques can also be applied to explore the set of sequential quantum correlations for the case of time-dependent operations and sequences of length three.

\section*{Note added} After completing this work, we have become aware of the related result in Ref.~\cite[]{wagner2018device}. In Ref.~\cite[]{wagner2018device} the authors discuss self-testing of binary unsharp qubit measurements in the framework of the Bell test. In Appendix~\ref{app:comp_fr} we give a quick comparison of the obtained results and the used frameworks. We would also like to draw the reader's attention to work in Ref.~\cite[]{mohan2019sequential}. 

\begin{acknowledgments} 
N.M.~would like to thank Costantino Budroni,  Nicolas Brunner, Mariami Gachechiladze, and Anubhav Chaturvedi for useful discussions. The work on this project was supported by First TEAM Grant No. 2016-1/5. The SDP optimization was performed with SDPT3 solver~\cite[]{toh1999sdpt3} and YALMIP toolbox~\cite[]{lofberg2004yalmip} for MATLAB.  
\end{acknowledgments} 

\begin{appendix}
\onecolumngrid
\section*{Appendix}

In this Appendix we provide technical details omitted in the main text. In Appendix~\ref{app:instr_rank_1} we discuss the optimality of L\"uders instruments, a fact regarding extremal instruments that we use in the main text. In Appendix~\ref{app:pvm_bound} we give a proof of Proposition~\ref{prop:pvm_bound}. Appendix~\ref{app:robustness} is dedicated to a more detailed description of our results on the robustness of SDI characterization of unsharp measurements. Apart from that, in Appendix~\ref{app:dephase} we derive an upper bound on the average success probability of $2\to 1$ QRAC for the case of a dephasing channel being applied on the states of Alice. These results are used for the robustness proofs. In Appendix~\ref{app:biased} we present our observations regarding the possibility of SDI characterization of biased two-outcome qubit measurements. Finally, in Appendix~\ref{app:comp_fr} we provide a comparison of our results with Ref.~\cite{wagner2018device}. \\

We start by fixing the notations that we use in this Appendix. We denote the preparation states of Alice by $\rho_\vx$, where $\vx\in \{(00),(01),(10),(11)\}$. The POVMs of Charlie we denote by $(C^z_0,C^z_1)$. We use the notation $\ket{\cdot^\perp}$ to denote the qubit state orthogonal to the given one, i.e. $\bk{\psi}{\psi^\perp}=0,\; \forall \ket{\psi} \in \Hl$, which is not ambiguous in the qubit case. We also use the common shorthand notation $\ket{i}\otimes\ket{i}=\ket{ii}$, as well as $\ket{+} = \frac{1}{\sqrt{2}}(\ket{0}+\ket{1})$,  $\ket{-} = \frac{1}{\sqrt{2}}(\ket{0}-\ket{1})$.

\section{Optimality of L\"uders instruments}
\label{app:instr_rank_1}
In this Appendix we show that the optimal instruments of Bob are of L\"uders type, which is essentially equivalent to saying that the optimal instruments have rank-$1$ 
Choi-Jamio\l{}kowski (CJ) operators. Using the formula for conditional probability from Eq.~\eqref{eq:p_distr} we can write explicitly the expression for the average success probability~\eqref{eq:f_merit} in terms of operators $\sigma^y_{\vx,b} = \mathcal{B}^y_b(\rho_\vx)$ as follows 
\beq
\Ps = \frac{1}{16}\sum_{\vx,y,z,b,c}\tr(\sigma^y_{\vx,b}C^z_c)(\alpha\delta_{b,x_y}+(1-\alpha)\delta_{c,x_z}).
\label{eq:psucc_sigma} 
\eeq
Let us introduce new operators $\rho^{y}_{\vx,\oplus} = \sigma^y_{\vx,0}\oplus\sigma^y_{\vx,1}$, $\rho^y_{\vx,\oplus} \in \Ll(\Hl'), \;\forall \vx,y$, with $\text{dim}(\Hl') = 4$. We can think of $\rho^{y}_{\vx,\oplus}$ as a block-diagonal matrix for each $y$ and $\vx$, with two blocks being $\sigma^y_{\vx,0}$ and $\sigma^y_{\vx,1}$. In a similar fashion we can construct operators $C^{y,z}_{\vx,c} = C^z_c(\alpha\delta_{0,x_y}+(1-\alpha)\delta_{c,x_z})\oplus C^z_c(\alpha\delta_{1,x_y}+(1-\alpha)\delta_{c,x_z})$,  $C^{y,z}_{\vx,c}\in \Ll(\Hl'), \;\forall \vx,y,z,c$. The expression in Eq.~\eqref{eq:psucc_sigma} can then be written simply as
\beq
\Ps = \frac{1}{16}\sum_{\vx,y,z,c}\tr(\rho^y_{\vx,\oplus}C^{y,z}_{\vx,c}).
\eeq  
 It is easy to see that each $\rho^{y}_{\vx,\oplus}$ is a state, i.e.,~$\rho^{y}_{\vx,\oplus} \geq 0$, $\tr(\rho^{y}_{\vx,\oplus}) = 1$. It is also clear that the set of $\rho^{y}_{\vx,\oplus}$ is convex and closed. Moreover, for the given POVMs of Bob $B^{y}_b$ we can always construct a ``measure-and-prepare" instrument of the form $\Bl^y_b(\cdot) = \tr(B^y_b\cdot)\tilde{\rho}^y_b$, where $\tilde{\rho}^y_b$ are states. The states $\rho^{y}_{\vx,\oplus}$ can then be expressed as $\rho^{y}_{\vx,\oplus} = \tr(\rho_\vx B^y_0)\tilde{\rho}^y_0\oplus \tr(\rho_\vx B^y_1)\tilde{\rho}^y_1$. Since $\tilde{\rho}^y_b$ can be chosen arbitrarily, we can conclude that the extreme elements of the set of $\rho^{y}_{\vx,\oplus}$ are direct sums of pairs of pure states, multiplied by $\tr(\rho_\vx B^y_0)$ and $\tr(\rho_\vx B^y_1)$. The proof can be obtained trivially by assuming the opposite and writing the spectral decomposition for each of $\tilde{\rho}^y_b$, which is then nothing but a convex combination of rank-1 block-diagonal operators (which means that the corresponding states are pure). Of course, measure-and-prepare instruments will not be optimal, but it is sufficient to consider this type of instrument to prove the optimality of rank-1 operators $\sigma^y_{\vx,b}$.

Now we can show that from the optimality of pure states of Alice $\rho_\vx$ and rank-1 operators $\sigma^y_{\vx,b}$ it follows that CJ operators of optimal Bob instruments are necessarily rank $1$. Let us write the spectral decomposition of $\mathfrak{B}^y_b$ for each $y,b$,
\beq
\label{eq:instr_sp}
\mathfrak{B}^y_b = \sum\limits_{k=0}^{3}\kb{\beta_k}{\beta_k},
\eeq       
where $\ket{\beta_k}$ are not normalized. We did not write the subscripts $y,b$ for $\ket{\beta_k}$ to avoid unnecessary complication, but, of course, $\ket{\beta_k}$ will be different for different $y,b$. Each $\ket{\beta_k}$ can be written in its Schmidt decomposition~\cite[]{schmidt1989theorie} as $\ket{\beta_k} = \ket{p_k}\otimes\ket{q_k}+\ket{p^\perp_k}\otimes\ket{q^\perp_k}$, where we can require $\ket{q_k},\ket{q^\perp_k}$ to be normalized, and leave $\ket{p_k},\ket{p^\perp_k}$ not normalized. If we denote $\rho_\vx = \kb{\psi_\vx}{\psi_\vx}$, we can write the following: 
\beq
\sigma^y_{\vx,b} = \sum_{k=0}^3\kb{q_{\vx,k}}{q_{\vx,k}}, \quad \text{with}\quad \ket{q_{\vx,k}} = \bk{\psi_\vx}{p_k}\ket{q_k}+\bk{\psi_\vx}{p^\perp_k}\ket{q^\perp_k}.
\eeq
However, we have already concluded that optimal $\sigma^y_{\vx,b}$ must be rank-1 operators, which means that all $\ket{q_{\vx,k}}$ must coincide up to a phase, i.e.,~$\ket{q_{\vx,k}} = e^{i\phi_{k,k'}}\ket{q_{\vx,k'}}$ for some phases $\phi_{k,k'}$. Now we can use the argument that the optimal $\ket{\psi_\vx}$ must necessarily be linearly independent, i.e.,~$\bk{\psi_\vx}{\psi_{\vx'}}\neq 1$ for at least two $\vx,\vx'$, $\vx\neq \vx'$. It is pretty clear that in the opposite case Alice is not sending any information to Bob. Since $\{\ket{\psi_\vx}\}_{\vec{x}}$ span the entire Hilbert space, we come to the conclusion that $\ket{p_k}\otimes\ket{q_k}+\ket{p^\perp_k}\otimes\ket{q^\perp_k} =  e^{\phi_{k,k'}}(\ket{p_{k'}}\otimes\ket{q_{k'}}+\ket{p^\perp_{k'}}\otimes\ket{q^\perp_{k'}})$, $\forall k\neq k'$, i.e., $\ket{\beta_k} = e^{\phi_{k,k'}}\ket{\beta_{k'}}$, $\forall k \neq k'$. However, since Eq.~(\ref{eq:instr_sp}) is a spectral decomposition, it must be that $\bk{\beta_k}{\beta_{k'}} = \delta_{k,k'}$, $\forall k \neq k'$, from which it follows that there is only one vector in the sum in Eq.~\eqref{eq:instr_sp}. This completes the proof. We should mention that the above proof can be directly extended to the higher-dimensional case. 
	
\section{Proof of Proposition~\ref{prop:pvm_bound}}
\label{app:pvm_bound}
In this Appendix we prove the bounds given in Proposition~\ref{prop:pvm_bound} on the average success probability in Eq.~\eqref{eq:f_merit} for the case of projective strategies of Bob. 
There are two types of projective measurements that we need to consider, $(\Id,0)$ and $(\proj{p},\proj{p^\perp})$. The case $(0,\Id)$ is clearly equivalent to $(\Id,0)$. Since we need to consider only instruments of Bob corresponding to rank-1 CJ operators $\mathfrak{B}^y_b = \proj{\beta^y_b}$, the latter can be taken of the form 
\beq
\label{eq:u_intr}
\ket{\beta^y_0} = \ket{0}\otimes\ket{q^y} + \ket{1}\otimes\ket{{q^y}^\perp}, \quad \ket{\beta^y_1} = 0,
\eeq 
for POVM $(\Id,0)$. In the main text of the paper we refer to this strategy as ``unitary" -- because the Kraus operator for this instrument is a unitary. Another possibility for $\mathfrak{B}^y_b = \proj{\beta^y_b}$ is the following:
\beq
\label{eq:mp_intr}
\ket{\beta^y_0} = \ket{p^y}\otimes\ket{q^y_0}, \quad \ket{\beta^y_1} = \ket{{p^y}^\perp}\otimes\ket{q^y_1},
\eeq 
which corresponds to POVM $(\proj{p^y},\proj{{p^y}^\perp})$. We refer to this strategy as ``measure and prepare". In both cases all the vectors $\ket{p^y}, \ket{q^y}, \ket{q^y_b}$, $y=0,1, b=0,1$ are normalized. It is important to note that in Eq.~\eqref{eq:u_intr} the relative phase between $\ket{q^y}$ and $\ket{{q^y}^\perp}$ should be taken into account. 

Due to the same requirement on the rank of $\mathfrak{B}^y_b$ we can directly see that the probabilistic strategies in which Bob applies, let us say, the operation from Eq.~\eqref{eq:u_intr} with some probability and the strategy from Eq.~\eqref{eq:mp_intr} in all other cases will necessarily be less optimal than one of these strategies chosen with probability $1$. It means that in order to analyze the bounds on the average success probability for strategies of Bob that correspond to projective measurements and their probabilistic mixtures it is sufficient to compute the bounds for the extreme projective strategies. Thus, we need to analyze the following three cases: (a) Bob applies a unitary strategy for $y=0,1$, (b) Bob applies measure-and-prepare strategy for all $y=0,1$, and (c) Bob applies a unitary strategy for $y=0$ and a measure-and-prepare strategy for $y=1$. In the main part of the text we refer to the last case as a ``mixed" strategy. It is also clear that strategies, equivalent to the ones listed, e.g., unitary strategy with the outcome $b$ flipped, or the mixed strategy with $y$ flipped, will give the same bounds. In what follows we  analyze in detail all three cases (a), (b), and (c). When combined, it gives the proof of Proposition~\ref{prop:pvm_bound}.

\subsection*{(a) ``Unitary" strategy} 
As mentioned above, the instrument from Eq.~\eqref{eq:u_intr} has a single Kraus operator that is a unitary. Let us denote it as $U_y = \kb{0}{q^y}+\kb{1}{{q^y}^\perp}$, $y=0,1$. Let us now write the expression for average success probability in terms of $U_y$: 
\beq
\Ps = \frac{\alpha}{8}\sum_{\vx,y}\delta_{x_y,0}+\frac{1-\alpha}{16}\sum_{\vx,y,z}\tr(U_y\rho_\vx U_y^\dagger C^z_{x_z}).
\eeq
It is clear that the first sum is equal to $4$. The second sum can be upper-bounded by $2\max_y(\sum_{\vx,z}\tr(U_y\rho_\vx U_y^\dagger C^z_{x_z}))$ and for the maximal $y$ we can introduce new $\tilde{C^z_c} =  U_y^\dagger C^z_c U_y$. Maximization over $\tilde{C^z_c}$ and $\rho_\vx$ is a standard optimization problem for $2 \to 1$ QRAC, and thus the upper bound on $\Ps$ in this case is $\frac{\alpha}{2} + (1-\alpha)(\frac{1}{2}+\frac{1}{2\sqrt{2}})$.

\subsection*{b) ``Measure and prepare" strategy} 
In the proof we use two main facts regarding the optimality of certain relations between states and measurements. The first one concerns maximization of expectation values $\tr(\rho(a\proj{\psi} + b\proj{\phi}))$ over states $\rho \in \Ll(\Hl)$, for some $a,b\in \mathbb{R}$, and $\ket{\psi},\ket{\phi}$ being normalized vectors in $\Hl$, $\dim(\Hl) = 2$. The maximum is obtained when $\rho=\proj{\xi}$ with $\ket{\xi}$ being a normalized eigenvector of $a\proj{\psi} + b\proj{\phi}$ corresponding to its maximal eigenvalue. When there are no further constraints on $\rho$, the maximum can always be attained and the following formula gives its explicit form:

\beq
\label{eq:norm}
\norm{a\proj{\psi} + b\proj{\phi}} = \frac{a+b}{2}+\frac{1}{2}\sqrt{(a-b)^2+4ab|\bk{\psi}{\phi}|^2},
\eeq
where we used the notation $\norm{\cdot}$ for the operator norm (operator's maximal eigenvalue).

The second fact concerns optimization of expressions of type $\tr(A(\proj{\psi}-\proj{\phi}))$ over $\ket{\psi}$ and $\ket{\phi}$, with both being normalized vectors in $\Hl$, with $\dim(\Hl) = 2$ and where $A \in \Ll(\Hl)$ is some linear, not necessarily positive, operator. It can be seen that for optimal $\ket{\psi}$ and $\ket{\phi}$ necessarily $\bk{\psi}{\phi} = 0$. To see that, let us write the spectral decomposition of the operator $\proj{\psi}-\proj{\phi} = \lambda\proj{\xi} - \lambda\proj{\xi^\perp}$, with $\bk{\xi}{\xi^\perp} = 0$, and $\lambda \in [-1,1]$. Then, $\max_{\ket{\psi},\ket{\phi}}(\tr(A(\proj{\psi}-\proj{\phi}))) = \max_{\ket{\xi}}(\max_{\lambda}(\lambda\tr(A(\proj{\xi}-\proj{\xi^\perp}))))$. Clearly, the maximum over $\lambda$ is attained at the boundary, i.e., for $\lambda \in \{1,-1\}$. It is also clear that for $\lambda = -1$, a simple relabeling $\xi^\perp \leftrightarrow \xi$ gives the same expression as for $\lambda = 1$. Without loss of generality we can take $\ket{\psi} = \ket{\xi}$ and $\ket{\phi} = \ket{\xi^\perp}$, which proves our claim.

Let us now proceed to the actual derivation of the bound. We remind ourselves that the CJ operators of the instruments of Bob have the following form (Eq.~\eqref{eq:mp_intr})
\beq
\ket{\beta^y_0} = \ket{p^y}\otimes\ket{q^y_0}, \quad \ket{\beta^y_1} = \ket{{p^y}^\perp}\otimes\ket{q^y_1},\; y=0,1.
\eeq 
After removing dependency due to POVM normalization, we can rewrite the average success probability as follows
\beq
\label{eq:app_pvm_instr}
\Ps = &&\frac{1}{2}+\frac{\alpha}{8}(r^0_0+r^0_1+r^1_0-r^1_1)+\frac{1-\alpha}{16}\Big((r^0_0+r^0_1)\tr((\proj{q^0_0}-\proj{q^0_1})C_0^0)+(r^0_0-r^0_1)\tr((\proj{q^0_0}-\proj{q^0_1})C^1_0)\nonumber \\
&&+(r^1_0+r^1_1)\tr((\proj{q^1_0}-\proj{q^1_1})C_0^0)+(r^1_0-r^1_1)\tr((\proj{q^1_0}-\proj{q^1_1})C_0^1)\Big),
\eeq
where $r^y_0 = \bra{p^y}\rho^T_{00}-\rho^T_{11}\ket{p^y}$ and $r^y_1 = \bra{p^y}\rho^T_{01}-\rho^T_{10}\ket{p^y}$. Now we can use the fact that the optimal states $\rho_\vx$ are pure. Here we take $\rho^T_\vx = \proj{\psi_\vx}$ and, from what we have discussed in the beginning of this section, we can assume $\bk{\psi_{00}}{\psi_{11}} = \bk{\psi_{01}}{\psi_{10}} = \bk{q^0_0}{q^0_1} = \bk{q^1_0}{q^1_1} = 0$. This creates some redundancy in Eq.~\eqref{eq:app_pvm_instr}, which can be removed. As a result one obtains the following expression for the optimal $\Ps$:
\beq
\label{eq:app_pvm_instr_2}
\Ps = &&\frac{3}{4}-\frac{\alpha}{2}+\frac{2\alpha-1}{4}(|\bk{p^0}{\psi_{00}}|^2+|\bk{p^1}{\psi_{00}}|^2)+\frac{\alpha}{4}(|\bk{p^0}{\psi_{01}}|^2-|\bk{p^1}{\psi_{01}}|^2)\nonumber \\ 
&&+\frac{1-\alpha}{4}\sum_{y=0}^1\bra{q^y_0}\Big(\proj{c_0^0}(|\bk{p^y}{\psi_{00}}|^2+|\bk{p^y}{\psi_{01}}|^2-1)+\proj{c_0^1}(|\bk{p^y}{\psi_{00}}|^2-|\bk{p^y}{\psi_{01}}|^2)\Big)\ket{q^y_0},
\eeq
where we have grouped the terms with respect to $\ket{q^y_0}$. We have also used the fact that extremal effects $C_0^z$ in this case are necessarily rank $1$ and introduced the notation $C_0^z = \proj{c_0^z}$. Applying Eq.~\eqref{eq:norm} to Eq.~\eqref{eq:app_pvm_instr_2} for the expression enclosed in $\ket{q^y_0}$ we obtain the following: 
\beq
\label{eq:app_pvm_instr_final}
\Ps = &&\frac{2-\alpha}{4}+\frac{\alpha}{4}\sum_{y=0}^1(|\bk{p^y}{\psi_{00}}|^2+(-1)^y|\bk{p^y}{\psi_{01}}|^2)\nonumber \\
&&+\frac{1-\alpha}{8}\sum_{y=0}^1\sqrt{(1-2|\bk{p^y}{\psi_{01}}|^2)^2+4(|\bk{p^y}{\psi_{00}}|^2-|\bk{p^y}{\psi_{01}}|^2)(|\bk{p^y}{\psi_{00}}|^2+|\bk{p^y}{\psi_{01}}|^2-1)|\bk{c_0^0}{c_0^1}|^2}.
\eeq
Each of the square roots are monotonic functions of $|\bk{c_0^0}{c_0^1}|^2$ and their domain is $[0,1]$, if all possible values of $|\bk{p^y}{\xi_i}|$ are considered. Depending on the signs of $(|\bk{p^y}{\psi_{00}}|^2-|\bk{p^y}{\psi_{01}}|^2)(|\bk{p^y}{\psi_{00}}|^2+|\bk{p^y}{\psi_{01}}|^2-1)$ the bounds on each square root will be obtained for $|\bk{c_0^0}{c_0^1}|=0$ or $|\bk{c_0^0}{c_0^1}|=1$. One can expect that this is not the case for the sum of the square roots, i.e.,~the total $\Ps$. However, one can upper-bound the maximum of sums of the square roots by the sum of the maxima. Thus, we need to consider two cases of $|\bk{c_0^0}{c_0^1}|=0,1$ for each of the square roots, i.e., four cases in total. By doing so we obtain the following:
\beq
&&\Ps \leq \frac{1}{4}\max(1+|\bk{p^0}{\psi_{00}}|^2+\alpha|\bk{p^0}{\psi_{01}}|^2+|\bk{p^1}{\psi_{00}}|^2-\alpha|\bk{p^1}{\psi_{01}}|^2, \nonumber\\
&&2-\alpha+|\bk{p^0}{\psi_{00}}|^2+\alpha|\bk{p^0}{\psi_{01}}|^2+\alpha|\bk{p^1}{\psi_{00}}|^2-|\bk{p^1}{\psi_{01}}|^2,\\
&&1+\alpha|\bk{p^0}{\psi_{00}}|^2+|\bk{p^0}{\psi_{01}}|^2+|\bk{p^1}{\psi_{00}}|^2-\alpha|\bk{p^1}{\psi_{01}}|^2,\nonumber \\
&&2-\alpha+\alpha|\bk{p^0}{\psi_{00}}|^2+|\bk{p^0}{\psi_{01}}|^2+\alpha|\bk{p^1}{\psi_{00}}|^2-|\bk{p^1}{\psi_{01}}|^2).
\eeq
For each of the four cases we can group the terms with respect to $\ket{p^y}$ and upper-bound each of these inner products by the norms of the respective operators that are expressions of $\proj{\psi_{00}}$ and $\proj{\psi_{01}}$ with coefficients depending on $\alpha$. It is interesting that in each of these four cases the final upper bound is $1/2+\sqrt{1+\alpha^2}/4$, which completes the proof.

\subsection*{(c) ``Mixed" strategy} 
In this proof we use similar techniques as in the previous case of the measure-and-prepare strategy. In particular, we use the formula from Eq.~\eqref{eq:norm} and, whenever our success probability will be written as an affine function of operators of the form $\proj{\psi}-\proj{\phi}$, we assume $\bk{\psi}{\phi} = 0$.

For the mixed strategy of Bob the formula for the average success probability takes the following form
\beq
\label{eq:app_mixed_1}
\Ps = &&\frac{\alpha}{8}\sum_\vx(\delta_{0,x_0})+\frac{\alpha}{4}(\tr((\rho^T_{00}+\rho^T_{10})\proj{p^1})+\tr((\rho^T_{01}+\rho^T_{11})\proj{{p^1}^\perp}))+ \frac{1-\alpha}{16}\sum_{\vx,z}\tr(U_0\rho_\vx U^\dagger_0C^z_{x_z})\nonumber \\
&&+\frac{1-\alpha}{16}\sum_{\vx,z}(\tr(\rho^T_\vx\proj{p^1})\tr(\proj{q^1_0}C^z_{x_z})+\tr(\rho^T_\vx\proj{{p^1}^\perp})\tr(\proj{q^1_1}C^z_{x_z})),
\eeq
where we have introduced the notation $U_0 = \kb{0}{q^0}+\kb{1}{{q^0}^\perp}$. Since we optimize the expression in Eq.~\eqref{eq:app_mixed_1} over $C^z_c$ and $\proj{q^1_b}$ and we have no restrictions on the basis in which we consider these operators, we can apply $U^\dagger_0$ to the basis of $C^z_c$ and $U_0$ to the basis of $\proj{q^1_b}$. Clearly, this will not change the optimal value for $\Ps$, but it removes this unitary from the expression in Eq.~\eqref{eq:app_mixed_1}. We again use normalization of all POVMs in Eq.~\eqref{eq:app_mixed_1} to simplify it, which gives the following expression:
\beq
\Ps = &&\frac{1}{2}+\frac{\alpha}{8}\bra{p^1}(r_0-r_1)\ket{p^1}+\frac{1-\alpha}{16}\Big(\bra{c^0_0}(r_0+r_1)\ket{c^0_0}+\bra{c^1_0}(r_0-r_1)\ket{c^1_0}\nonumber \\
&&+\bra{p^1}(r_0+r_1)\ket{p^1}\bra{c^0_0}(\proj{q^1_0}-\proj{q^1_1})\ket{c^0_0}+\bra{p^1}(r_0-r_1)\ket{p^1}\bra{c^1_0}(\proj{q^1_0}-\proj{q^1_1})\ket{c^1_0}\Big),
\eeq    
where we used the fact that the optimal $C^z_0 = \proj{c^z_0}$ is rank $1$ and used a slightly different notation for $r_0, r_1$, which are in this case: $r_0 = \proj{\psi_{00}}-\proj{\psi_{11}}$, $r_1 = \proj{\psi_{01}}-\proj{\psi_{10}}$, with $\rho^T_\vx = \proj{\psi_\vx}$ as before. Now we can make use of the orthogonality relations $\bk{\psi_{00}}{\psi_{11}}=\bk{\psi_{01}}{\psi_{10}} = \bk{q^1_0}{q^1_1}=0$ to obtain the following:
\beq
\label{eq:app_mixed_2}
\Ps = &&\frac{2-\alpha}{4}+\frac{2\alpha-1}{4}|\bk{p^1}{\psi_{00}}|^2+\frac{\alpha}{4}|\bk{p^1}{\psi_{10}}|^2 + \frac{1-\alpha}{8}(|\bk{c^0_0}{\psi_{00}}|^2-|\bk{c^0_0}{\psi_{10}}|^2+|\bk{c^1_0}{\psi_{00}}|^2+|\bk{c^1_0}{\psi_{10}}|^2)\nonumber\\
&&+\frac{1-\alpha}{4}\Big(|\bk{q^1_0}{c^0_0}|^2(|\bk{p^1}{\psi_{00}}|^2-|\bk{p^1}{\psi_{10}}|^2)+|\bk{q^1_0}{c^1_0}|^2(|\bk{p^1}{\psi_{00}}|^2+|\bk{p^1}{\psi_{10}}|^2-1)\Big). 
\eeq
From this, it was not obvious how to proceed with the proof, so we tried to maximize the expression in Eq.~\eqref{eq:app_mixed_2} numerically. We used a simple parametrization of each vector with two real parameters and performed an unconstrained heuristic optimization over $12$ real parameters with gradient method for each value of $\alpha$. From this numerical optimization we could infer that the maximum of $\Ps$ is obtained when $|\bk{q^1_0}{c^0_0}| = 1/2$ and $|\bk{q^1_0}{c^1_0}|=1$. We prove the validity of these observations later, but for now we assume them true and finish the calculations. 

Taking $|\bk{q^1_0}{c^0_0}| = 1/2$ and $|\bk{q^1_0}{c^1_0}|=1$ reduces Eq.~\eqref{eq:app_mixed_2} to the following form: 
\beq
\label{eq:app_mixed_3}
\Ps = &&\frac{1}{4} + \frac{1+\alpha}{8}\bra{p^1}(\proj{\psi_{00}}+\proj{\psi_{10}})\ket{p^1}+ \frac{1-\alpha}{8}\bra{c^1_0}(\proj{\psi_{00}}+\proj{\psi_{10}})\ket{c^1_0}\nonumber \\
&&+\frac{1-\alpha}{8}\bra{c^0_0}(\proj{\psi_{00}}-\proj{\psi_{10}})\ket{c^0_0}.
\eeq
Now we can put an upper bound on the expression in Eq.~\eqref{eq:app_mixed_3} by substituting all expectation values of operators $\proj{\psi_{00}}\pm\proj{\psi_{10}}$ by their norms. Using Eq.~\eqref{eq:norm} we write $\Ps$ as a function of $|\bk{\psi_{00}}{\psi_{10}}|$, which is $\Ps = \frac{1}{2} + \frac{|\bk{\psi_{00}}{\psi_{10}}|}{4} + \frac{1-\alpha}{8}\sqrt{1-|\bk{\psi_{00}}{\psi_{10}}|^2}$. Finally, optimization over $|\bk{\psi_{00}}{\psi_{10}}|$ gives the result stated in the proposition, i.e., $\Ps \leq \frac{1}{2}+\frac{\sqrt{4+(1-\alpha)^2}}{8}$.

Now, let us verify the assumptions $|\bk{q^1_0}{c^0_0}| = 1/2$ and $|\bk{q^1_0}{c^1_0}|=1$. In order to do so, let us write all the states in Eq.~\eqref{eq:app_mixed_2} in Bloch vector form. Since no basis is fixed, we can always assume $\proj{\psi_{00}} = (\Id + \sigma_z)/2$. We can also assume, without loss of generality, that $\proj{\psi_{10}} = (\Id + x_1\sigma_x + z_1\sigma_z)/2$, i.e.,~that $\proj{\psi_{10}}$ lies in the $x$-$z$ plane of the Bloch ball. The normalization condition for $x_1,z_1$ reads $x^2_1 + z^2_1 = 1$. For the rest of the states, in general, we will have all $x,y,z$ components. Let us take $\proj{p^1} = (\Id + x_2\sigma_x + y_2\sigma_2 + z_2\sigma_z)/2$, $\proj{q^1_0} = (\Id + x_3\sigma_x + y_3\sigma_2 + z_3\sigma_z)/2$, $\proj{c^0_0} = (\Id + x_4\sigma_x + y_4\sigma_2 + z_4\sigma_z)/2$, and $\proj{c^1_0} = (\Id + x_5\sigma_x + y_5\sigma_2 + z_5\sigma_z)/2$. The normalization condition states that $x^2_i + y^2_i + z^2_i = 1$, for all $i=2,3,4,5$. Inserting this form of states to Eq.~\eqref{eq:app_mixed_2} gives the following expression:
\beq
\label{eq:app_mixed_4}
\Ps = &&\frac{1}{2} + \frac{\alpha}{8}(z_2+x_1x_2+z_1z_2) + \frac{1-\alpha}{16}(z_4+z_5+x_1x_5+z_1z_5-x_1x_4-z_1z_4) + \frac{1-\alpha}{16}z_2(x_3x_4+y_3y_4+z_3z_4 \nonumber \\
&&+ x_3x_5+y_3y_5+z_3z_5) - \frac{1-\alpha}{16}(x_1x_2+z_1z_2)(x_3x_4+y_3y_4+z_3z_4-x_3x_5-y_3y_5-z_3z_5).
\eeq 
We can show now that for solution maximizing the expression above we have necessarily $y_2=y_3=y_4=y_5=0$. In order to do so, let us consider the optimization problem
\beq
\label{eq:app_mixed_sub}
\max &\quad & z_4+z_5+x_1x_5+z_1z_5-x_1x_4-z_1z_4 \\
\text{s.t.} &\quad & x_3^2+y_3^2+z_3^2 = 1,\nonumber\\
 & & x_4^2+y_4^2+z_4^2 = 1,\nonumber\\
 & & x_5^2+y_5^2+z_5^2 = 1,\nonumber\\
 & & x_4x_3 + y_4y_3 + z_4z_3 = v_1,\nonumber\\
 & & x_5x_3 + y_5y_3 + z_5z_3 = v_2,\nonumber
\eeq 
where maximization is carried over vectors $(x_3,y_3,z_3), (x_4,y_4,z_4)$, and $(x_5,y_5,z_5)$. For the fixed vectors $(x_1,0,z_1), (x_2,y_2,z_2)$ and the inner products $v_1,v_2\in [-1,1]$, this optimization problem is equivalent to the original maximization of the expression in Eq.~\eqref{eq:app_mixed_4}. Let us now construct a Lagrangian of the problem in Eq.~\eqref{eq:app_mixed_sub},
\beq
\label{eq:app_mixed_sub_lag}
\Ll = &&z_4+z_5+x_1x_5+z_1z_5-x_1x_4-z_1z_4 + \sum_{i=1}^3\lambda_i(x_i^2+y_i^2+z_i^2 - 1)\nonumber \\
&&+\lambda_4(x_4x_3 + y_4y_3 + z_4z_3 - v_1) + \lambda_5(x_5x_3 + y_5y_3 + z_5z_3 - v_2),
\eeq
where $\lambda_i, i\in\{1,\dots,5\}$ are the Lagrange multipliers. The system of equations for stationary points of Eq.~\eqref{eq:app_mixed_sub_lag} contains, among others, the following three systems of linear equations
\beq
&&\begin{pmatrix} 2\lambda_1 & \lambda_4 & \lambda_5 \\ \lambda_4 & 2\lambda_2 & 0 \\  \lambda_5 & 0 & 2\lambda_3 \end{pmatrix}
\begin{pmatrix} y_3 \\ y_4 \\ y_5 \end{pmatrix} = \begin{pmatrix} 0 \\ 0 \\ 0 \end{pmatrix}, \label{eq:app_mixed_s1}\\
&&\begin{pmatrix} 2\lambda_1 & \lambda_4 & \lambda_5 \\ \lambda_4 & 2\lambda_2 & 0 \\  \lambda_5 & 0 & 2\lambda_3 \end{pmatrix}
\begin{pmatrix} x_3 \\ x_4 \\ x_5 \end{pmatrix} = \begin{pmatrix} 0 \\ x_1 \\ -x_1 \end{pmatrix}, \label{eq:app_mixed_s2}\\
&&\begin{pmatrix} 2\lambda_1 & \lambda_4 & \lambda_5 \\ \lambda_4 & 2\lambda_2 & 0 \\  \lambda_5 & 0 & 2\lambda_3 \end{pmatrix}
\begin{pmatrix} z_3 \\ z_4 \\ z_5 \end{pmatrix} = \begin{pmatrix} 0 \\ 1+z_1 \\ 1-z_1 \end{pmatrix}. \label{eq:app_mixed_s3}
\eeq
We can immediately notice that the first system in Eq.~\eqref{eq:app_mixed_s1} has nontrivial solutions if and only if the determinant of the system's matrix is $0$. However, if this is the case, and $x_1\neq 0, z_1\neq \pm 1$, the second two systems have no solutions. Since we know that the optimal solution to the problem in Eq.~\eqref{eq:app_mixed_sub} has to be among the stationary points, we can conclude that necessarily $y_3=y_4=y_5=0$. Now we just need to check the exceptions, which are the cases when $x_1\neq 0, z_1\neq \pm 1$. It is easy to see that in these cases the objective function becomes either $2z_4$ or $2z_5$, i.e.,~the maximum attained in the problem in Eq.~\eqref{eq:app_mixed_sub} is at most $2$. However, we know from our numerical solution that the optimal value of the expression $z_4+z_5+x_1x_5+z_1z_5-x_1x_4-z_1z_4$ is greater than $2$ for all $\alpha<1$. 

Proving the optimality of $y_2=0$ is even easier. We can simply remove it from consideration by substituting the constraint $x^2_2 + y^2_2 + z^2_2 = 1$ with $x^2_2 + z^2_2 \leq 1$. Since the expression in Eq.~\eqref{eq:app_mixed_4}, which we need to maximize, is linear in $x_2$ and $z_2$, it means that the maximum is attained at the boundary of the subspace of $x_2$ and $z_2$, i.e., when $x^2_2 + z^2_2 = 1$. This means in practice that $y_2=0$.  

We can apply use the SDP relaxation of Lasserre to the polynomial optimization problems~\cite[]{lasserre2006convergent}. We considered the second level of the Lasserre hierarchy in which case the size of the moment matrix $\Gamma$ is $66$ and the number of variables in the dual SDP is exactly $1000$. An important note, which might be of interest to those who would like to reproduce the result or solve a similar problem, is that in the optimization it was crucial to add the normalization constraints of the type $\Gamma(x_i,x_ix_j)+\Gamma(z_i,z_ix_j) = \Gamma(1,x_j)$ and $\Gamma(x_i,x_iz_j)+\Gamma(z_i,z_iz_j) = \Gamma(1,z_j)$, $i,j\in \{1,\dots,5\}$ to the obvious constraints $\Gamma(x_i,x_i)+\Gamma(z_i,z_i) = 1$, $i\in \{1,\dots,5\}$. If the former are not included in the SDP, the objective function in Eq.~\eqref{eq:app_mixed_4} turns out to be unbounded. 

By performing the above SDP optimization we were able to confirm the optimality of the conditions $|\bk{q^1_0}{c^0_0}| = 1/2$ and $|\bk{q^1_0}{c^1_0}|=1$ up to numerical precision of $10^{-10}$ as well as reproduce the value of $1/2+\sqrt{4+(1-\alpha)^2}/8$ of the bound up to the same precision. This completes our proof.

\section{Robustness of self-testing}
\label{app:robustness}

In this Appendix we give more details concerning the robustness of our SDI characterization of unsharp measurements. The question that we would need to answer is the following: Given the experimentally obtained value of the average success probability for some $\alpha$, $\Ps^{\text{exp}} \leq \Ps^{\text{POVM}}(\alpha)$, what can we say about the unsharpness parameter of the implemented measurements? Here we again discuss the case of trace-1 POVMs of Bob. In order to answer the posed question we first try to answer the opposite question: Given the eigenvalues of the POVMs of Bob, what is the maximal average success probability that can be obtained? 

We start by fixing the CJ operator of the first element of Bob's instrument to be 
\beq
\ket{\beta^0_0} = \sqrt{\lambda_0}\ket{00}+\sqrt{1-\lambda_0}\ket{11},
\eeq  
where now $\lambda_0$ is the operator norm of the effects of the first POVM (its largest eigenvalue). We can then apply the seesaw method of Ref.~\cite{werner2001bell} in order to establish the optimal form of other instruments of Bob and measurements of Charlie. From these numerical results we infer that we can make the following assumptions:
\beq
\label{eq:rob_expl}
&&C^0_0 = \proj{0},\quad C^0_1 = \proj{1},\quad C^1_0 = \proj{+},\quad C^1_1 = \proj{-}, \nonumber \\
&&\ket{\beta^0_1} = \sqrt{1-\lambda_0}\ket{00}+\sqrt{\lambda_0}\ket{11} \\
&&\ket{\beta^1_0} = \sqrt{\lambda_1}\ket{++}+\sqrt{1-\lambda_1}\ket{--},\quad\ket{\beta^1_1} = \sqrt{1-\lambda_1}\ket{++}+\sqrt{\lambda_1}\ket{--},\nonumber
\eeq
where now $\lambda_1$ can be different from $\lambda_0$. We confirm later that these choices of Charlie's measurements and Bob's instruments are indeed optimal, but for now we assume that it is true and continue with the derivations of the bound.  

Using Eq.~\eqref{eq:rob_expl} we can obtain the bound on the average success probability for given $\lambda_0$ and $\lambda_1$. The states $\rho_\vx$ can be chosen to be the eigenstates of the respective operators corresponding to the operators' maximal eigenvalues, as it is often done in such proofs. We get the following
\beq
\label{eq:rob_psucc}
&&\Ps(\lambda_0,\lambda_1) = \frac{1}{2}+\frac{1}{4}\sqrt{\frac{(1-3\alpha)^2}{2}+F(\lambda_0,\lambda_1)},\; \text{where}\nonumber\\
&&F(\lambda_0,\lambda_1) = (1-\alpha)(1-3\alpha)(\sqrt{\lambda_1(1-\lambda_1)}+\sqrt{\lambda_0(1-\lambda_0)})-(1+\alpha)(1-3\alpha)(\lambda_0^2+\lambda_1^2)\\
&&+4\alpha(1-\alpha)(\sqrt{\lambda_0(1-\lambda_0)}\lambda_1+\lambda_0\sqrt{\lambda_1(1-\lambda_1)})+(1-5\alpha^2)(\lambda_0+\lambda_1).\nonumber
\eeq
If we fix $\lambda_0$ to be some value between $\frac{1}{2}$ and $1$, and optimize the expression in Eq.~\eqref{eq:rob_psucc} w.r.t.~$\lambda_1$, we find that the optimal $\lambda_1$ is in fact the one given by Eq.~\eqref{eq:opt_norm_app}. The same goes for the optimal $\lambda_0$, if $\lambda_1$ is fixed. It means that, if we want to find a rectangular area in the space of $(\lambda_0,\lambda_1)$ for which $\Ps$ from Eq.~\eqref{eq:rob_psucc} is greater than or equal to some observed value $\Ps^{\text{exp}}$, we should fix one of the norms, let us say $\lambda_0$, to be one given by Eq.~\eqref{eq:opt_norm_app}, and solve the inequality $\Ps(\lambda_0,\lambda_1) \geq \Ps^{\text{exp}}$ with respect to $\lambda_1$. Unfortunately, solution to this inequality is too unwieldy to be written explicitly, even for this Appendix. However, since it is a single-variable problem it can be solved with numerical methods up to an arbitrary precision. It is instructive to give the solution in graphical form, which we do in the main text (see Fig.~\ref{fig:norm}).     

Let us now discuss the optimality of the solution in Eq.~\eqref{eq:rob_psucc}. We further modified the SDP techniques, described in Section~\ref{sec:method} in the main text, in order to account for the fixed values of $\lambda_0$ and $\lambda_1$. We consider the vectors $\ket{\beta^y_b}$ to be of the following form:
\beq
\label{eq:app_schmidt}
&&\ket{\beta^y_0} = \sqrt{\lambda_y}\ket{p^y}\otimes\ket{q^y_0}+\sqrt{1-\lambda_y}\ket{{p^y}^\perp}\otimes\ket{{q^y_0}^\perp}\nonumber\\
&&\ket{\beta^y_1} = \sqrt{1-\lambda_y}\ket{p^y}\otimes\ket{q^y_1}+\sqrt{\lambda_y}\ket{{p^y}^\perp}\otimes\ket{{q^y_1}^\perp},
\eeq
which is the general form of the vectors of CJ operators of instruments corresponding to two unbiased POVMs with effects' norms $\lambda_0$ and $\lambda_1$. 

As a first modification, we add two operators $\{\proj{p^y}\otimes\Id\}_{y=0,1}$ to the set of operators $\mathcal{O}$, where $\ket{p^y}$ are meant to be the vectors in the Schmidt decompositions from Eq.~\eqref{eq:app_schmidt}. Considering these operators allows us to add the following constraints to Eq.~\eqref{eq:sdp}:
\beq
\label{eq:add_c_1}
&&\Gamma^y_0(\proj{p^y}\otimes\Id,\Id\otimes\Id) = \lambda_y,\; \Gamma^y_1(\proj{p^y}\otimes\Id,\Id\otimes\Id) = 1-\lambda_y,\; y=0,1,\nonumber\\
&&\Gamma^y_0(\rho_\vx\otimes\Id,\Id\otimes\Id) = 1-\lambda_y+\frac{2\lambda_y-1}{\lambda_y}\Gamma^y_0(\rho_\vx\otimes\Id,\proj{p^y}\otimes\Id),\; y=0,1,\\
&&(1-\lambda_y)\Gamma^y_0(\rho_\vx\otimes\Id,\proj{p^y}\otimes\Id) = \lambda_y\Gamma^y_1(\rho_\vx\otimes\Id,\proj{p^y}\otimes\Id),\; y=0,1.\nonumber
\eeq
On top of that, we set the upper bound on the success probability of Charlie, if dephasing channels with parameters $\frac{1}{2}-\sqrt{\lambda_0(1-\lambda_0)},\frac{1}{2}-\sqrt{\lambda_1(1-\lambda_1)}$ are applied to the states of Alice. In particular, we enforce the following:
\beq
\label{eq:add_c_2}
\sum_{\vx,z,c,b}\delta_{x_z,c}\Big((-1)^c\Gamma^y_b(\rho_\vx\otimes\Id,\Id\otimes C^z_0)+\delta_{c,1}\Gamma^y_b(\rho_\vx\otimes\Id,\Id\otimes\Id)\Big)\leq 4+2\sqrt{1+4\lambda_y(1-\lambda_y)}. 
\eeq 
After adding the constraints from Eqs.~(\ref{eq:add_c_1}) and (\ref{eq:add_c_2}) to the SDP in Eq.~\eqref{eq:sdp} we can compare the bounds with the predictions given in Eq.~\eqref{eq:rob_psucc}. Since we know that for deriving the bounds on the norms we need to take one of $\lambda_y$, let us say $\lambda_0$, to be of the optimal form in Eq.~\eqref{eq:opt_norm_app}, we just need to make a comparison for each pair $(\alpha,\lambda_1)$. What we found, is that the modified SDP gives the exact bound from Eq.~\eqref{eq:opt_norm_app} up to the numerical precision ($\sim 10^{-10}$), whenever $\lambda_1\leq\lambda_0$, i.e.,~for all $\lambda_1$ less than or equal to the optimal norm. For $\lambda_1>\lambda_0$ we obtained a slight discrepancy in the bounds. 

We believe that the optimal solution is of the form in Eq.~\eqref{eq:rob_expl} and the proposed bound is tight. However, for the results to be rigorous, we need to estimate the ``error" in the estimation of the norms due to the fact that we could not reproduce exactly the bounds with the SDP method. In order to do so, we can use the simple error estimation formula $\Delta\lambda_1 = \frac{\partial\lambda_1}{\partial\Ps(\lambda_0,\lambda_1)}\Delta\Ps$. We have included the estimated ``errors" in Fig.~\ref{fig:norm} as crosses of the respective colors corresponding to different relative discrepancies in the observed average success probability. 

As a final remark we give a proof of the bound used in Eq.~\eqref{eq:add_c_2}. First of all, we can notice that for the channel corresponding to one of the settings of Bob we can take $\ket{q^y_0} = \ket{q^y_1}$. This can be easily seen if we write down the CJ operator of the channel:
\beq
\label{eq:app_channel}
&&\sum_{b=0,1}\proj{\beta^y_b} = \proj{p^y}\otimes(\lambda_y\proj{q^y_0}+(1-\lambda_y)\proj{q^y_1})+\proj{{p^y}^\perp}\otimes((1-\lambda_y)\proj{{q^y_0}^\perp}+\lambda_y\proj{{q^y_1}^\perp})\nonumber\\
&&+\sqrt{\lambda_y(1-\lambda_y)}\Big(\kb{p^y}{{p^y}^\perp}\otimes(\kb{q^y_0}{{q^y_0}^\perp}+\kb{q^y_1}{{q^y_1}^\perp})+\kb{{p^y}^\perp}{p^y}\otimes(\kb{{q^y_0}^\perp}{q^y_0}+\kb{{q^y_1}^\perp}{q^y_1})\Big).
\eeq
We can now rewrite this expression taking into account the normalization conditions, which yields the following:
\beq
&&\sum_{b=0,1}\proj{\beta^y_b} = \Id\otimes\Id-\Id\otimes\Big((1-\lambda_y)\proj{q^y_0}+\lambda_y\proj{q^y_1}\Big)-\proj{p^y}\otimes\Id+\proj{p^y}\otimes(\proj{q^y_0}+\proj{q^y_1})\nonumber\\
&&+\sqrt{\lambda_y(1-\lambda_y)}\Big(\kb{p^y}{{p^y}^\perp}\otimes(\kb{q^y_0}{{q^y_0}^\perp}+\kb{q^y_1}{{q^y_1}^\perp})+\kb{{p^y}^\perp}{p^y}\otimes(\kb{{q^y_0}^\perp}{q^y_0}+\kb{{q^y_1}^\perp}{q^y_1})\Big).
\eeq

We are interested in finding the maximum of the following average success probability of Charlie:
\beq
\label{eq:psucc_ch}
\Ps = \frac{1}{8}\sum_{\vx,z}\tr(\rho^T_\vx\otimes C^z_{x_z} \sum_{b=0,1}\proj{\beta^y_b}).
\eeq
We can easily see that $\sum_{\vx,z}\tr\Big(\rho^T_\vx\otimes C^z_{x_z} \Id\otimes\Big((1-\lambda_y)\proj{q^y_0}+\lambda_y\proj{q^y_1}\Big)\Big) = 4$. From here it follows that the resulting expression is a linear function in $\proj{q^y_0}+\proj{q^y_1}$ and $\kb{q^y_0}{{q^y_0}^\perp}+\kb{q^y_1}{{q^y_1}^\perp}$. Suppose now the optimal $\ket{q^y_0}\neq \ket{q^y_1}$ and one of the vectors gives a larger contribution to the value of the expression in Eq.~\eqref{eq:psucc_ch}. In this case taking both vectors equal to the optimal one will give a larger value of $\Ps$, which contradicts our assumptions. It can also happen that $\ket{q^y_0}\neq \ket{q^y_1}$ and the values of the expressions corresponding to both $\ket{q^y_0},\ket{q^y_1}$ are equal. In that case, however, there is no difference in taking them equal or not. 

Let us now consider the channel
\beq
\proj{p^y}\otimes\proj{q^y_0}+\proj{{p^y}^\perp}\otimes\proj{{q^y_0}^\perp}+2\sqrt{\lambda_y(1-\lambda_y)}\Big(\kb{p^y}{{p^y}^\perp}\otimes\kb{q^y_0}{{q^y_0}^\perp}+\kb{{p^y}^\perp}{p^y}\otimes\kb{{q^y_0}^\perp}{q^y_0}\Big),
\eeq
which is the same channel as in Eq.~\eqref{eq:app_channel}, but with $\ket{q^y_1}=\ket{q^y_0}$.
We now derive the bound for each value of $y$ independently. Since we have a freedom of fixing bases in this case, we can set $\ket{p^y} = \ket{q^y_0} = \ket{0}$. We can then see that the channel we are considering is nothing but a dephasing channel of the form $\Lambda(\rho) = (1-p)\rho + p\sigma_z\rho\sigma_z$ with parameter $p=\frac{1}{2}-\sqrt{\lambda_y(1-\lambda_y)}$. We derive the bound in the following Appendix.

\subsection*{Success probability of $2\to 1$ QRAC under a dephasing channel}
\label{app:dephase}
Let us consider a $2\to 1$ QRAC with two parties Alice and Charlie, Alice's preparation states being $\{\rho_\vx\}_{\vx}$ and Charlie's measurements $\{C^z_c\}_{c=0,1}$. Let us assume that the states of Alice undergo a transformation, corresponding to a dephasing channel $\Lambda(\rho) = (1-p)\rho + p\sigma_z\rho\sigma_z$ with some parameter $p\in[0,\frac{1}{2}]$. The average success probability which we would like to maximize is the following expression:
\beq
\label{eq:app_pscucc_qrac}
\Ps = \frac{1}{8}\sum_{\vx,z}\tr(\Lambda(\rho_\vx)C^z_{x_z}).
\eeq
We now prove that $\Ps\leq \frac{1}{2}+\frac{1}{4}\sqrt{1+(1-2p)^2}$.
First, let us expand the expression in Eq.~\eqref{eq:app_pscucc_qrac} and use the normalization condition for POVMs of Charlie. We obtain the following:
\beq
\Ps = \frac{1}{8}\tr\Big(\Lambda(\rho_{01}+\rho_{10}+2\rho_{11})+\Lambda(\rho_{00}-\rho_{11})(C^0_0+C^1_0)+\Lambda(\rho_{01}-\rho_{10})(C^0_0-C^1_0)\Big).
\eeq
Now we use the argument that we made already several times in this paper, that for optimal states of Alice we have $\rho_{01}+\rho_{10}=\Id$ and $\rho_{00}+\rho_{11}=\Id$. This leaves us with the following function to maximize,
\beq
\Ps = \frac{1}{4}+\frac{1}{4}\tr\Big(\Lambda(\rho_{00})(C^0_0+C^1_0)+\Lambda(\rho_{01})(C^0_0-C^1_0)\Big),
\eeq
where we used the fact that the dephasing channel is unital. Now let us write the states and POVM effects in Bloch form introducing the following notation:
\beq
&&\rho_{00} = \frac{1}{2}(\Id+x_0\sigma_x+y_0\sigma_y+z_0\sigma_z),\; \rho_{01} = \frac{1}{2}(\Id+x_1\sigma_x+y_1\sigma_y+z_1\sigma_z),\nonumber\\
&&C^0_0 = \frac{1}{2}(\Id+x_2\sigma_x+y_2\sigma_y+z_2\sigma_z),\; C^1_0 = \frac{1}{2}(\Id+x_3\sigma_x+y_3\sigma_y+z_3\sigma_z).
\eeq
It is easy to see that the action of the dephasing channel is in multiplying the variables $x_0,x_1,y_0$, and $y_1$ by the factor $(1-2p)$. The expression for $\Ps$ in terms of the Bloch coefficients reads
\beq
\Ps = \frac{1}{2}+\frac{1}{8}\Big(z_0(z_2+z_3)+z_1(z_2-z_3)+(1-2p)\Big(x_0(x_2+x_3)+y_0(y_2+y_3)+x_1(x_2-x_3)+y_1(x_2-x_3)\Big)\Big).
\eeq
Without loss of generality we can take $x_0=y_0=0$, $z_0=1$. We are then left with the following optimization problem
\beq
\max &\quad & \frac{1}{2}+\frac{1}{8}\Big(z_2+z_3+z_1(z_2-z_3)+(1-2p)(x_1(x_2-x_3)+y_1(y_2-y_3))\Big)\\
\text{s.t.} &\quad & x_i^2+y_i^2+z_i^2=1,\; i=1,2,3.\nonumber
\eeq
This problem can be easily solved using the Lagrange multipliers method, giving the bound of $\frac{1}{2}+\frac{1}{4}\sqrt{1+(1-2p)^2}$.

\section{``Biased" case}
\label{app:biased}
Here we discuss the ``biased" case, i.e.,~the case when $\tr(B^y_b)\neq 1$, $b=0,1, y=0,1$. We use $(\lambda_0,\lambda_1)$ to denote the spectrum of the first effects of Bob's POVMs. The second effects then have the spectrum $(1-\lambda_0,1-\lambda_1)$. These, in principle, can depend on $y$, however, in our case they do not. 

	\begin{figure}[h!] \centering
		\includegraphics[width=.7\textwidth]{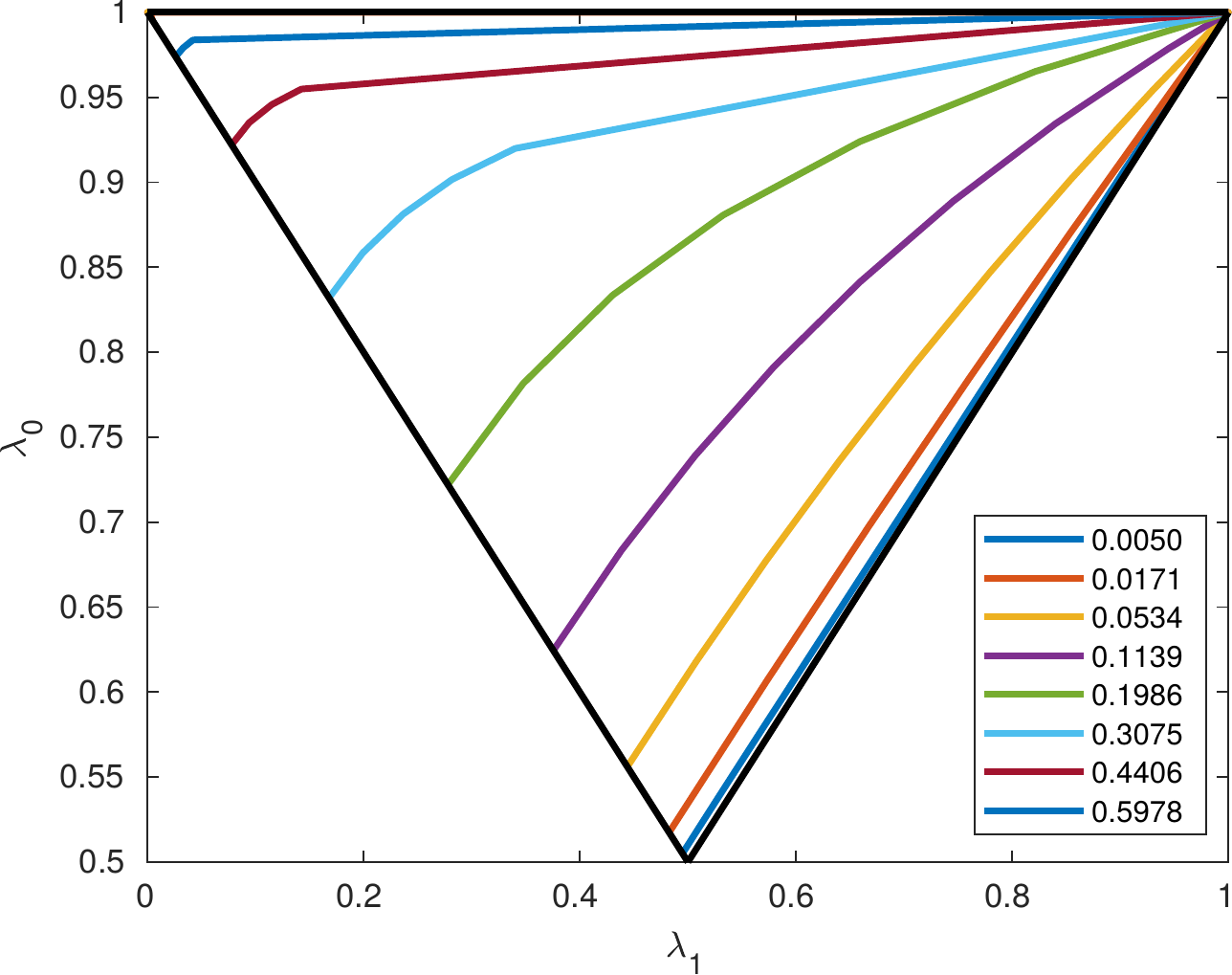}
		\caption{Plot for various $\lambda_0$ and $\lambda_1$. Different lines correspond to different values of $\alpha$. The lines are ordered from top to bottom on the plot as $\alpha$ changes from $0.0050$ to $0.5978$.}
		\label{fig:biased}
	\end{figure}

The figure of merit can be adjusted to account for the ``bias" and here we consider the following expression for the average success probability
\begin{equation}
\label{eq:f_merit_biased}
\Ps = \frac{\alpha}{2}\sum\limits_{\vx,y}p(\vx)\text{Pr}(b=x_y|\vx,y) + \frac{1-\alpha}{8}\sum\limits_{\vx,z} \text{Pr}(c=x_z|\vx,z),
\end{equation}
with $p(\vx)$ being the probability distribution $p(00) = \gamma^2, p(01)=p(10)=\gamma(1-\gamma), p(11) = (1-\gamma)^2$, with $\gamma\in [0, 1]$. Notice that we do not change the distribution of $\vx$ for Charlie. The case $\gamma=\frac{1}{2}$ corresponds to the original figure of merit in Eq.~\eqref{eq:f_merit}, with which we provided the SDI characterization of the ``unbiased" case, i.e.,~$\lambda_0+\lambda_1 = 1$.  For $\gamma\neq \frac{1}{2}$, however, we noticed that the optimal eigenvalues of the first effect of Bob's POVMs do not sum to $1$. We summarized our findings in Fig.~\ref{fig:biased}. 

Let us first discuss the values of $(\lambda_0,\lambda_1)$ that can occur. First of all, we know that $0\leq \lambda_i\leq 1$, $i=0,1$, since the second effects have to be positive. We can also consider the case when $\lambda_0+\lambda_1\geq 1$, since we have a freedom of denoting which effect is the first or the second. The corresponding region in space of $(\lambda_0,\lambda_1)$ is a triangle, that is depicted with black lines in Fig.~\ref{fig:biased}. 

The edge of this triangle that connects the vertices $(0.5,0.5)$ and $(0,1)$ corresponds to the ``unbiased" case of $\lambda_0+\lambda_1=1$. If we wish to self-test POVMs corresponding to this line we take $\gamma=\frac{1}{2}$ and vary $\alpha$ between $0$ and $1$, as discussed in the main text. If we would like to ``move" towards the vertex $(1,1)$ of this triangle, we need to increase the parameter $\gamma$ from $\frac{1}{2}$ to $1$. In Fig.~\ref{fig:biased} we have depicted with color lines what happens with optimal values of $(\lambda_0,\lambda_1)$, for the given value of $\alpha$, if we increase the parameter $\gamma$. These are the results of the optimization with the seesaw method of Ref.~\cite{werner2001bell}. As we can see from these plots, we can cover all the area of the triangle by choosing appropriate $(\alpha,\gamma)$. This supports our claims of the fact that all binary measurements on a qubit can be characterized in the proposed scenario. The robustness analysis of this characterization should be reported elsewhere.

\section{Comparison between SDI framework and the one used in Ref.~\cite{wagner2018device}}
\label{app:comp_fr}

In Ref.~\cite{wagner2018device} the authors discuss self-testing of unbiased binary qubit measurements in the Bell scenario. Normally, self-testing in the Bell scenario is a stronger result than the similar one in the SDI framework, because one does not assume the dimension of the underlined system. We would like to argue that, first of all, our scheme can also be adapted to the device-independent framework considered in Ref.~\cite{wagner2018device}. Second, we consider the SDI framework to be much more natural for self-testing of unsharp measurements. 

Below we propose a general, quite intuitive way to map SDI scenarios to the one considered in Ref.~\cite{wagner2018device}. Let us say we are interested in optimizing a linear function of probabilities $p(b|\vx,y) = \tr(\rho_\vx B^y_b)$, where $\rho_\vx \in \Ll(\Hl), \forall \vx$ are preparation states and $B^y_b\in \Ll(\Hl), \forall y,b$ are POVM effects of the measuring party. In the SDI framework we assume that $\dim(\Hl) \leq d$, where $d\geq2$. At this point we do not restrict the number of outcomes or the number of settings. Since we can always represent two sequential measurements as a single POVM, the following reasoning also applies to the scenario considered in the main text of this paper. 

Let us now consider the same linear function as before, but now we substitute the probabilities with $p(0,b|\vx,y) = \tr(A^\vx_0\otimes B^y_b \rho_{AB})$, where $A^\vx_0$ are the effects of POVMs of Alice corresponding to outcome $0$ and $\rho_{AB}$ is the state shared between the parties. If we can certify in a separate setup, as it is done in Ref.~\cite[]{wagner2018device}, that $\rho_{AB} = \proj{\Psi}$ is a maximally entangled state $\ket{\Psi} = \frac{1}{\sqrt{d}}\sum_{i=1}^d\ket{ii}$ with local dimension $d$ then $p(0,b|\vx,y) = \frac{1}{d}\tr((A^\vx_0)^T B^y_b)$. We can then relabel $(A^\vx_0)^T=\rho_\vx$ and return to the original optimization problem (up to a global factor $\frac{1}{d}$). This simple reasoning suggests that the case of maximal violation of the CHSH inequality in the scheme of Ref.~\cite{wagner2018device} is equivalent to the assumption on the dimension in the SDI framework.

We would also like to argue that our scheme is the simplest and also more natural for self-testing of unsharp measurements. Indeed, as clearly stated in the main text, unsharp measurements cannot be differentiated from their probabilistic realizations with PVMs in a simple two-party scenario, where the information about the post-measurement states is discarded. This concerns both the prepare-and-measure and the Bell-type scenario considered in Ref.~\cite[]{wagner2018device}. We would like to argue that the SDI framework, corresponding to the prepare-and-measure scenario is much more natural, because there is simply no way to avoid the assumption on the dimension of the Hilbert space of states that Bob sends to Charlie. If we lift this assumption, one can think about Bob and Charlie as a single party. In this case it is not possible to distinguish between two sequential binary measurements and the four-outcome POVM that one obtains by applying the Heisenberg dual of Bob's instruments to POVMs of Charlie. This assumption, although not explicitly stated, is also present in Ref.~\cite{wagner2018device}. From our perspective, if one has to make the assumption on the dimension of the Hilbert space in which $\rho^{y}_{\vx,b}$ are defined, there is no benefit in trying to relax this assumption for the states $\rho_\vx$ of Alice.

\end{appendix}
\twocolumngrid

\bibliographystyle{ieeetr}
\bibliography{two_rec_povm}
	
\end{document}